\documentclass[11pt]{article}

\usepackage[final]{acl}

\usepackage{times}
\usepackage{latexsym}

\usepackage[T1]{fontenc}

\usepackage[utf8]{inputenc}

\usepackage{microtype}

\usepackage{inconsolata}

\usepackage{graphicx}

\usepackage{xurl}

\usepackage{amsmath}
\usepackage{booktabs}
\usepackage{xspace}
\usepackage{xcolor}
\usepackage[normalem]{ulem}
\usepackage{makecell}
\usepackage[most]{tcolorbox}
\usepackage{graphicx}
\usepackage{subcaption}
\usepackage{enumerate}
\usepackage{multirow}
\usepackage[htt]{hyphenat}  %
\usepackage[title]{appendix}
\usepackage{cleveref}
\usepackage{fontawesome5}

\usepackage{balance}
\usepackage{listings}
\usepackage{xcolor}
\lstdefinelanguage{json}{
  basicstyle=\ttfamily\small,
  showstringspaces=false,
  breaklines=true,
  frame=single,
  backgroundcolor=\color{gray!5},
  stringstyle=\color{blue},
  keywordstyle=\color{purple},
}

\definecolor{mypink}{RGB}{233,30,99}

\newcommand{\xhdr}[1]{{\vspace{1mm}\noindent{\textbf{#1}}}}

\newcommand{\lloom}[0]{LLooM\xspace}

\newcommand{\wildchat}[0]{WildChat\xspace}
\newcommand{\allsides}[0]{AllSides\xspace}
\newcommand{\regactions}[0]{Regulatory Actions\xspace}
\newcommand{\msmarco}[0]{MS Marco\xspace}
\newcommand{\prods}[0]{Products\xspace}
\newcommand{\science}[0]{Science\xspace}

\newcommand{\organic}[0]{{\tt Organic}\xspace}
\newcommand{\aio}[0]{{\tt AIO}\xspace}
\newcommand{\gemini}[0]{{\tt Gemini}\xspace}
\newcommand{\gptsearch}[0]{{\tt GPT-Search}\xspace}  %
\newcommand{\gptnormal}[0]{{\tt GPT-Tool}\xspace}         %
\newcommand{\sonar}[0]{{\tt Sonar}\xspace}  %

\newcommand{\eg}[0]{\textit{e.g.,}\xspace}
\newcommand{\ie}[0]{\textit{i.e.,}\xspace}
\newcommand{\vs}[0]{\textit{v.s.}\xspace}

\title{Characterizing Web Search in the Age of Generative AI}

\author{
  \textbf{Elisabeth Kirsten\textsuperscript{1,3}},
 \textbf{Jost Grosse Perdekamp\textsuperscript{1,3}},
 \textbf{Qinyuan Wu\textsuperscript{2}},
 \textbf{Mihir Upadhyay\textsuperscript{1}}
\\
 \textbf{Krishna P. Gummadi\textsuperscript{2}},
 \textbf{Muhammad Bilal Zafar\textsuperscript{1,3}}
\\
 \textsuperscript{1}UA Ruhr Research Center for Trustworthy Data Science and Security,
 \\
 \textsuperscript{2}Max Planck Institute for Software Systems,
 \\
 \textsuperscript{3}Ruhr University Bochum
\\
 \small{
   \textbf{Correspondence:}
   \href{mailto:elisabeth.kirsten@rub.de}{\color{mypink}elisabeth.kirsten@rub.de}
 }
\\[-0.25em]
 \small{
 \faGithub\,{
 \href{https://github.com/aisoc-lab/generative-search-eval}{\color{mypink}\texttt{github.com/aisoc-lab/generative-search-eval}}}
 }
}

\begin{document}

\maketitle

\begin{abstract}
The advent of LLMs has given rise to \textit{generative search}, a new search paradigm in which LLMs retrieve information from the web related to a query and synthesize it into a single, coherent response. This paradigm differs fundamentally from traditional web search, where results are returned as a ranked list of independent web pages. In this paper, we ask: Along what dimensions does generative search differ from traditional search?
We conduct a systematic comparison between Google organic search and five generative search systems from three providers: Google, OpenAI, and Perplexity. Our analysis reveals substantial variation among engines in their reliance on internal \vs external knowledge, source diversity, and stability. While generative systems often achieve topical coverage comparable to traditional search, they do so using markedly different retrieval footprints and synthesis strategies. We further show that the outputs of generative search can vary across time and executions, raising new challenges for robustness. Our findings demonstrate that generative search introduces new dimensions that are not captured by existing evaluation paradigms, motivating the development of evaluations that explicitly account for retrieval behavior, synthesis, and stability in generative search systems.

\end{abstract}

\section{Introduction}
\label{sec:intro}

Search has been a mainstay of online information retrieval for three decades.
In response to a user query, traditional search engines 
return a ranked list of roughly $10$ web pages, ordered primarily by relevance and source authority~\cite{page1999pagerank}, but also influenced by factors like diversity, recency, and personalization~\cite{qinDiversifyingTopKResults2012}.

\begin{figure*}[t]
    \centering
    \begin{subfigure}[b]{0.32\textwidth}
        \includegraphics[width=\textwidth]{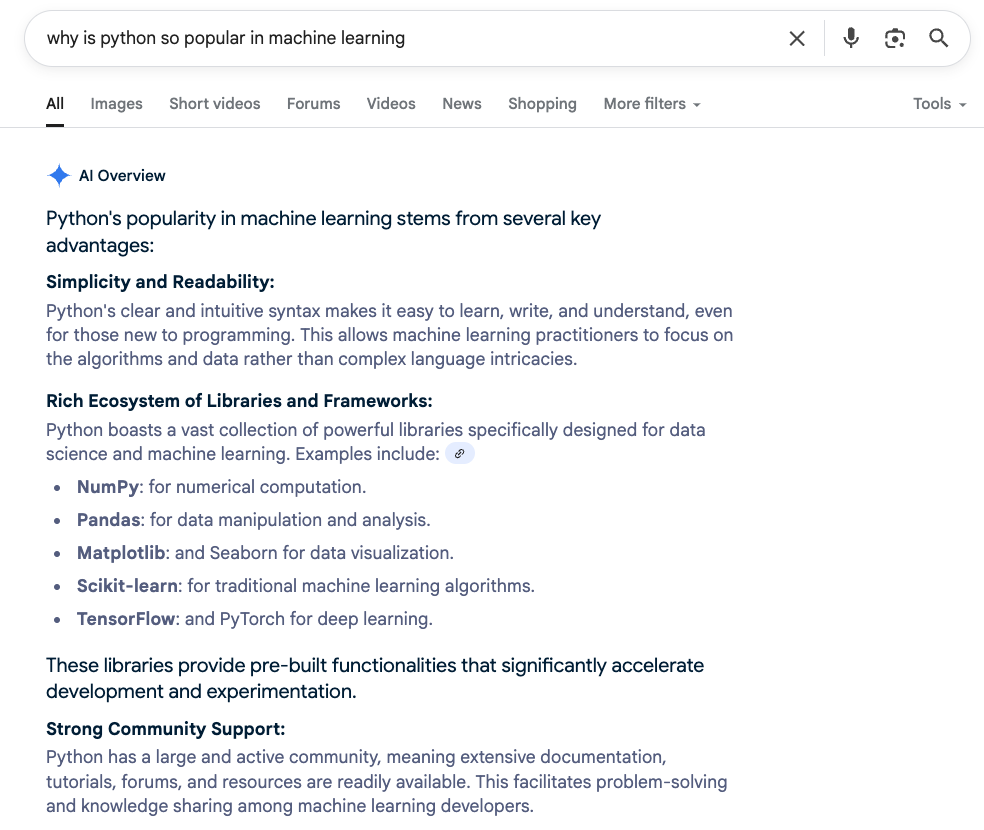}
        \caption{AI Overview}
        \label{fig:search_aio}
    \end{subfigure}
    \begin{subfigure}[b]{0.32\textwidth}
        \includegraphics[width=\textwidth]{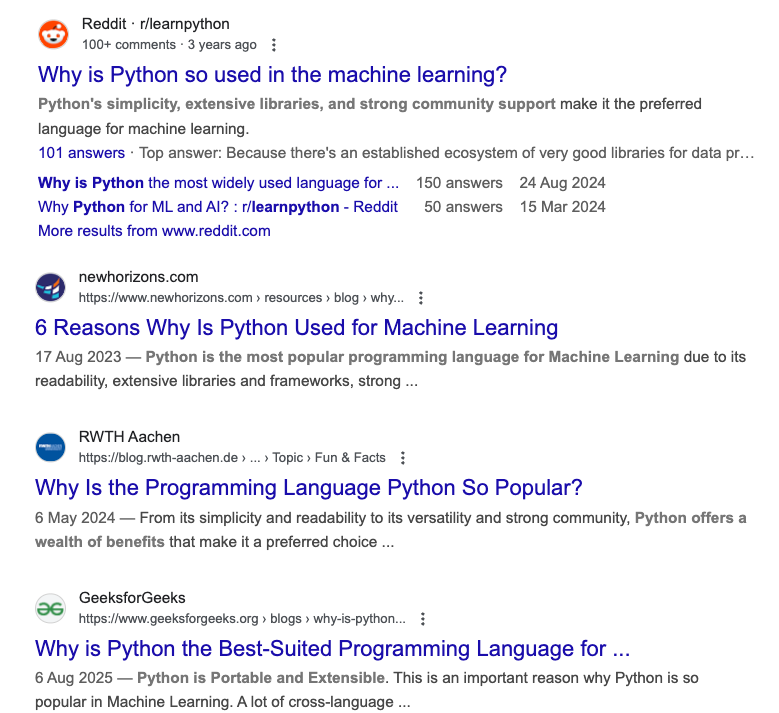}
        \caption{Organic search}
        \label{fig:search_organic}
    \end{subfigure}
    \begin{subfigure}[b]{0.32\textwidth}
        \includegraphics[width=\textwidth]{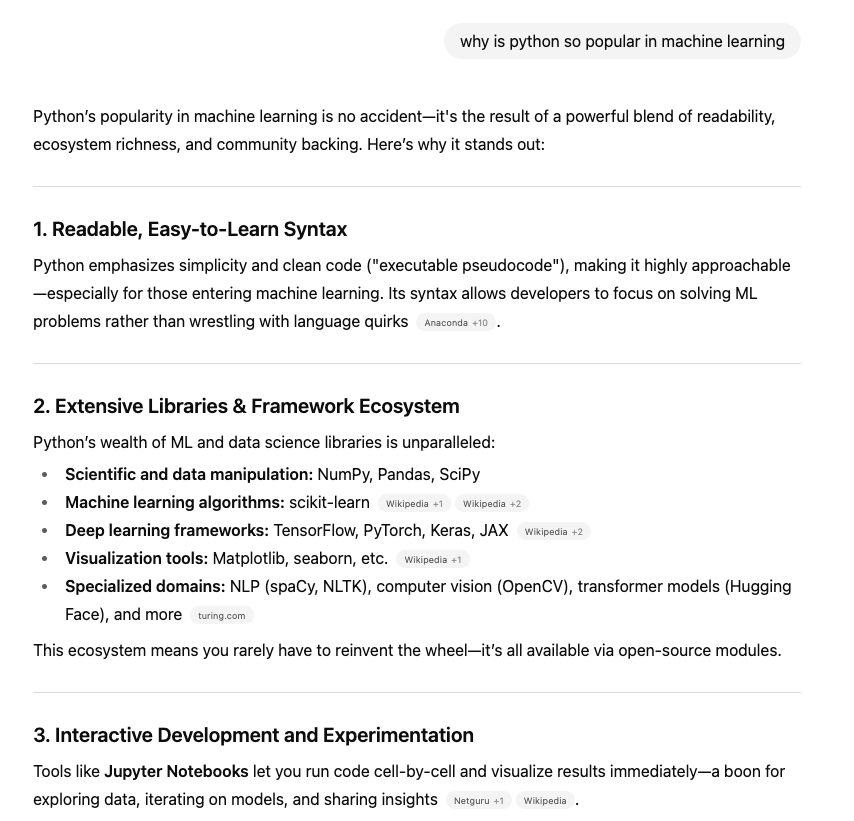}
        \caption{GPT-4o with search tool}
        \label{fig:search_gpt}
    \end{subfigure}
    \caption{
    Outputs of different search engines when querying for ``why is python so popular in machine learning''.
    \Cref{fig:search_aio,fig:search_organic} show the Google search output. At the top of the results page is a so-called ``AI Overview''.
    Below the AI Overview are the traditional web search results consisting of (often) top-10 web pages.
    The response of GPT-4o
    with web search used as a tool 
    is also accompanied by the supporting links (\Cref{fig:search_gpt}).
    }
    \label{fig:search_example}
\end{figure*}

The advent of LLMs has given rise to a new type of web search, generative AI-based search~\cite{liu2024llm,nakano2021webgpt}.
Under this new search paradigm, users 
receive answers in \textit{natural language rather than a ranked list of results}.
In fact, traditional search engines, including Google, now integrate generative search results in their outputs.
These systems typically operate by performing a web search, retrieving relevant pages, and producing a coherent, self-contained response.
See \Cref{fig:search_example} for an example.

While generative and traditional web search differ in many aspects, three fundamental dimensions stand out:
{\bf (i) Reliance on external \vs internal knowledge: }
Traditional web search operates by ranking external documents, \ie web pages. In contrast, generative search may rely solely on the underlying LLM's \textit{internal knowledge}, on \textit{external sources}, or on a combination of both~\cite{google_aio_rocks,nakano2021webgpt}.
{\bf (ii) Much wider coverage:} Traditional search engines typically present links and snippets from the top-10 web pages. Users, limited by their information processing capacity, would need to manually navigate in order to view the lower-ranked results. Users rarely inspect results beyond the top-10, and often not beyond the top-3~\cite{10.5555/1484611.1484615,stacy_search,urmanYouAreHow2023}.
Generative search, by contrast, can potentially aggregate information from tens of sources into a single response.
{\bf (iii) Synthesis of information using LLMs:}
Traditional search presents retrieved content verbatim as independent snippets (Figure~\ref{fig:search_organic}). 
In contrast, generative search synthesizes retrieved content into novel text. %
Because the underlying LLMs performing this synthesis are stochastic, issuing the same query multiple times can yield different outputs.

The impact of these fundamental differences on search outputs remains largely unexplored.
Most existing work on evaluating generative search focuses on properties such as factual accuracy, hallucinations, and bias (\S\ref{sec:related}), but does not explore, for instance, 
if generative search actually leverages its ability to go beyond top-10 web pages or if the stochasticity of the underlying LLM leads to materially different outputs across multiple runs.
To address this gap, we construct an evaluation framework that operationalizes these three key dimensions and uses them to systematically characterize the behavior of generative search systems.

We use our framework to compare a traditional search engine, Google's traditional search (\organic), with five generative search engines: Google AI Overview (\aio), Gemini (\gemini), GPT-4o-Search (\gptsearch), GPT-4o with search as a tool (\gptnormal), and Perplexity Sonar (\sonar) across a wide range of datasets covering domains like politics, shopping, and science.
Our evaluation reveals several intriguing patterns.

\xhdr{Generative  engines differ widely in how much they rely on internal \vs external knowledge.}
For the same queries, \gptnormal consults less than one web page on average, while \sonar, \aio, \gemini, and \gptsearch retrieve $14$, $9$, $9$, and $4$ on average, respectively.
Despite these differences in retrieval footprint, generative engines often achieve \textbf{similar levels of topical coverage}.
We also find that many generative engines perform web searches for retrieving simple, static facts that could be answered from internal knowledge alone (\eg \textit{What is the capital of France?}), lowering efficiency. %
However, engines that perform fewer searches sometimes produce inaccurate information on dynamic, time-sensitive queries (\eg \textit{seahawks vs steelers}).
This dichotomy reveals inherent \textbf{tradeoffs between efficiency and accuracy}.

Generative engines also differ in the breadth of sources they surface. %
They frequently \textbf{cite sources far beyond the top-10 or even top-100 organic search results.}
For instance, on average $53\%$ ($27\%$) of domains that \aio consults are not contained in top-10 (top-100) \organic search results.
Similarly, while $38\%$ ($89\%$) of the \organic result domains are contained within top-1K (top-1M) most visited websites, the corresponding numbers are $34\%$ ($85\%$) for \aio and $35\%$ ($81\%$) for \gptnormal.

Because generative engines rely on stochastic LLMs and complex retrieval pipelines, their \textbf{outputs are unstable across executions.}
For instance, when performing the same queries two months apart, only $18\%$ of web pages were common between the two runs of \aio, whereas the overlap was $45\%$ for \organic search.
Even over relatively short intervals of $5$ minutes ($24$ hours),
responses to questions requiring ternary answers (yes, no, neither) changed in up to $27\%$ ($28\%$) of cases.
Such instability raises concerns about reproducibility, trust, and user expectations, particularly when answers are presented as authoritative summaries.

Overall, our framework and findings show that generative search introduces new trade-offs in sourcing behavior, efficiency,  and stability that are not captured by existing search evaluation paradigms.
These results motivate the development of tailor-made evaluation methods that explicitly account for fundamental differences between the mechanics of traditional and generative search.

\section{Related Work}
\label{sec:related}

\xhdr{Evaluation of traditional web search} has long focused on issues like relevance, diversity, freshness, and coverage.
Relevance metrics such as Precision, Recall, and nDCG measure how well returned documents satisfy the user's information need and reward systems that rank relevant results highly.
Evaluation frameworks also measure \textit{diversity} to ensure coverage of multiple intents, subtopics, or viewpoints for ambiguous queries (\eg subtopic recall, $\alpha$-nDCG).
\textit{Freshness} captures the timeliness of results, particularly for event-driven queries, and \textit{coverage} measures the breadth of retrieved content \cite{lewandowski2012framework}.
These criteria are effective for ranked lists of documents, but do not directly apply to the synthesized, single-response outputs of generative engines.

\xhdr{Large Language Models (LLMs)} are commonly assessed on question answering, summarization, factual grounding, and tool use~\citep{liang2023holistic,pmlr-v235-huang24x,huang2024planningcreationusagebenchmarking}.
Summarization metrics capture coverage, coherence, and factual accuracy using measures such as ROUGE and BERTScore.
In information-seeking contexts, retrieval-augmented generation (RAG) enhances LLMs with external knowledge to improve factuality by incorporating external knowledge 
\citep{heZeroIndexingInternetSearch2024, SHI2025114354, joTaxonomyAnalysisSensitive2025}.
Recent work has also applied LLMs to query understanding, ranking \citep{sunChatGPTGoodSearch2024}, and query refinement \citep{siroAGENTCQAutomaticGeneration2024, bacciuGeneratingQueryRecommendations2024}.
In this work, we focus on a specific use case of LLMs as information sources in deployed, user-facing web search systems.

\xhdr{Prior work on diversity in search} has examined multiple dimensions of bias, including political, geographical, and commercial bias~\citep{jiangSearchConcentrationBias2014}.
Evaluations typically focus on the diversity of cited sources in the result set~\citep{kingraniDiversityAnalysisWeb2015,jiangSearchConcentrationBias2014,urmanAuditingSourceDiversity2021}, as well as ideological skew~\citep{linTrappedSearchBox2023}, and commercial bias~\citep{jiangSearchConcentrationBias2014}.
Recent work has applied ranking fairness metrics to quantify viewpoint diversity in search results~\citep{drawsAssessingViewpointDiversity2021} and measures coverage over predefined sets of perspectives~\citep{chenOpenWorldEvaluationRetrieving2025, skoutasIncreasingDiversityWeb, drawsAssessingViewpointDiversity2021}.
In this work, we measure diversity at both the source and content levels, and examine how retrieval and synthesis jointly shape the information presented to users.

\xhdr{Generative search} retrieves and synthesizes information into new text rather than returning a ranked list~\citep{SHI2025114354,nakano2021webgpt}.
Recent work evaluates generative search along dimensions such as verifiability, credibility, accuracy, and bias \citep{liuEvaluatingVerifiabilityGenerative2023,huEvaluatingRobustnessGenerative2024,daiBiasUnfairnessInformation2024,liGenerativeAISearch2024}. %
Other studies examine user interaction, trust, and feedback mechanisms in generative search interfaces \citep{aliannejadiInteractionsGenerativeInformation2024,daiNExTSearchRebuildingUser2025,sharmaGenerativeEchoChamber2024, Mayerhofer2025BlendingQA}. %
Recent work proposes new benchmarks, user models, and evaluation principles tailored to generative search \citep{narayananvenkitSearchEnginesAI2025, gienappEvaluatingGenerativeAd2024,aiInformationRetrievalMeets2023, alaofiGenerativeInformationRetrieval2025, miroyan2025searcharenaanalyzingsearchaugmented}
and analyzes when a search should be invoked~\citep{Schick2023ToolformerLMA, Li2025AdaptiveTUA,Sha2025SEMRLA}.
To the best of our knowledge, our work is the first to systematically characterize the sources that generative search engines retrieve and to study how differences in retrieval behavior affect the knowledge breadth, efficiency, and stability of generated content.

\section{Experimental Setup}
\label{sec:setup}

\begin{table}[t]
    \centering
    \resizebox{\columnwidth}{!}{
    \begin{tabular}{l|l|l}
        {\bf Dataset} & {\bf Domain} & {\bf Example queries} \\
        \midrule
        \msmarco 
        & \makecell{General\\(search engine)}
        & \makecell[l]{
            - origin of term doldrums \\ 
            - knowledge based technology definition \\ 
            - what causes typhoid fever
        }
        \\
        \hline
        \wildchat
        & \makecell{General\\(chatbot)}
        & \makecell[l]{
            - how do i stop procrastinating \\ 
            - what is rca in and in which part \\ \quad it is used in \\
            - which tech CEO is worth more than \\ \quad \$ 1 billion
        }
        \\
        \hline
        \allsides
        & \makecell{Politics}
        & \makecell[l]{
            - what is the personal income tax \\ 
            - what is terrorism in 100 words \\
            - how does the global economy \\ \quad affect jobs and career
        }
        \\
        \hline
        \makecell[l]{Regulatory\\Actions}
        & \makecell{Politics}
        & \makecell[l]{
            - what alternatives are offered \\ \quad  after ending DEI programs? \\
            - is my personal bitcoin affected by \\ \quad the strategic bitcoin reserve and stockpile?
        }
        \\
        \hline
        \makecell[l]{\science}
        & \makecell{Science}
        & \makecell[l]{
            - what is discrete search \\
            - what is set based programming \\
            - what company is leading in robotics
        }
        \\
        \hline
        \makecell[l]{\prods}
        & \makecell{Shopping}
        & \makecell[l]{
            - crocs worth it \\
            - school supplies reviews \\
            - best bedroom storage dresser
        }
        \\
        \hline
        \makecell[l]{Trends}
        & \makecell{Recency}
        & \makecell[l]{
            - when does ios 26 come out \\
            - emmy winners 2025 \\
            - ricky hatton cause of death
        }
    \end{tabular}
    }
    \caption{Example queries from each dataset. The datasets span five domains: politics, science, shopping, general real-world queries, and recent search trends.}
    \label{table:example_queries}
\end{table}

This section outlines the datasets and engines used.

\subsection{Datasets}
\label{sec:datasets}

Our datasets are designed to (i) reflect real user queries, (ii) cover both everyday and domain-specific workloads (\eg politics, science, products), and (iii) include both persistent and time-sensitive topics.
With these desiderata in mind, we consider the following datasets:

\xhdr{\msmarco.} The dataset contains real Bing search queries for open-domain retrieval and QA~\cite{bajaj2016ms}. We randomly sampled $1,000$ queries. 

\xhdr{\wildchat.} We subsample the dataset of 1 million queries released by \citet{zhao2024wildchat1mchatgptinteraction}. The authors gathered the dataset by granting users free access to ChatGPT and recording their interactions with the model. The authors of the original dataset post-processed it to remove PII, sensitive, and toxic conversations.
Since our goal is to compare search results of traditional and generative search engines, we exclude queries that are conversational in nature.
Specifically, we filter out queries that contain ``you'', ``your'' or ``u''.
We also filter out queries shorter than $4$ or longer than $75$ characters.
We only select queries that start with ``who'', ``what'', ``when'', ``where'', ``why'', ``how'', ``which'', sampling $250$ queries from each starting word.
In total, we obtain $1,750$ queries.

\xhdr{\allsides.} We generate this dataset to specifically focus on political issues. We take a list of political topics and issues from the news media site AllSides.com.%
\footnote{\url{https://www.allsides.com/topics-issues}}
We excluded topics if they were primarily geographic (\eg China, Russia), economic without direct political framing (\eg Business, Banking and Finance), cultural (\eg Arts and Entertainment), or overly broad (\eg World, General News) or ambiguous (Criminal Justice).
Examples of included topics are ``abortion'', ``immigration'' and ``religion and faith''.
We convert the topics to search queries as follows: We feed the topic to Google as a search query and gather 10 user-centric questions using Google's ``People Also Ask'' feature. We also include the topic itself as a query.

\xhdr{\regactions.} We generate this dataset to capture political queries pertaining to recent events. We gathered a list of major executive orders issues by the second Trump administration. We use the list maintained by the Brookings Institute.%
\footnote{\url{http://brookings.edu/articles/tracking-regulatory-changes-in-the-second-trump-administration/} (Accessed: 18/07/25)}
In total, we gathered 58 executive orders covering the timespan between January and July 2025.
For each action, we ask GPT-4o to generate 10 questions that a real person might ask about the action. The prompt is provided in \Cref{app:datasets}.

\xhdr{\science.} Our goal here was to gather queries related to scientific topics. 
We manually gathered 45 AI-related topics from the ACM Computing Classification System (CCS).%
\footnote{\url{https://dl.acm.org/ccs}}
Topics include entries like ``Information extraction'', ``Machine translation'' and ``Vagueness and fuzzy logic''.
We pass the topics to Google as a search query and gather 10 user-centric queries listed under the ``People Also Ask'' heading.

\xhdr{\prods.}
With this dataset, our goal was to study the difference in search results on product-related queries. 
We take a list of the 100 most searched Amazon products of 2023 from semrush.%
\footnote{\url{https://www.semrush.com/blog/most-searched-items-amazon/} (Accessed: 18/07/25)}
Searches include terms like ``apple watch band'', ``desk'' and ``iphone 13 case''.
We then turn the queries into review and comparison-oriented questions using custom templates like ``<product name> review'' and ``<product name> worth it'' (see \Cref{app:datasets} for details).

\xhdr{Trends.}
To analyze the behavior of different search engines in response to recent events, we create a dataset of trending queries.
We retrieve the top-100 search trends from Google Trends on September 15th, 2025, and query each search engine on the same day.
Google Trends\footnote{\url{https://trends.google.com}} analyzes the popularity of search terms in Google Search across the world, reflecting current public attention.

All the datasets combined consist of 4,706 queries.
\Cref{table:example_queries} shows examples of queries from each dataset.
Further details on dataset construction are provided in \Cref{app:datasets}.

\begin{figure*}[ht]
        \centering
        \includegraphics[width=\linewidth]{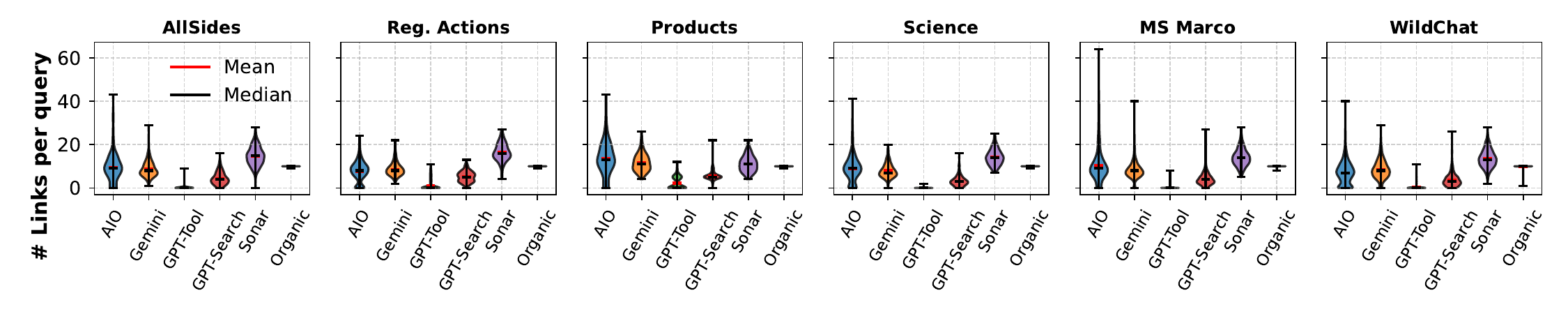}
    \caption{
    Different search engines rely on external knowledge to varying degrees.
    \gptnormal cites much fewer web pages than other engines, followed by \gptsearch. \aio tends to cite the most web pages.
    }
    \label{fig:num_links}
\end{figure*}

\subsection{Search Engines}
We compare one traditional and five generative search systems, selected to cover both dedicated search engines and general-purpose LLMs with search capabilities.
\textit{(i) Organic Google Search} (\organic) retrieves ranked lists of web pages. By default, the first page shows top-10 search results. In order to compare the overlap between generative search and Google search at various ranks, we retrieved top-100 search results.
\textit{(ii) Google AI Overviews} (\aio) generate synthesized answers together with cited sources. 
\textit{(iii) Gemini-2.5-Flash with Google Search} (\gemini) generates responses with an optional web search.
\textit{(iv) Perplexity Sonar} (\sonar) always performs a web search before returning an answer.
\textit{(v) GPT-4o Search} (\gptsearch) also always performs a web search.
\textit{(vi) GPT-4o with Search Tool} (\gptnormal) decides per query whether to retrieve external information.

\xhdr{Experimental Details.}
All queries were issued in English in September 2025. For generative engines, we set temperature to $0$ and maximum new tokens to $1,000$. Additional implementation details are in \Cref{app:experimental}.
To study the effect of location, we performed queries from two locations: United States ({\bf US}) and Germany ({\bf DE}). We observe only minor differences. Unless mentioned otherwise, the main paper reports results for US.
We discuss the results for the DE location in \Cref{app:de_analysis}.

\section{Internal \vs External Knowledge}
\label{sec:internal_vs_external}

We now characterize how different search engines balance internal knowledge and external information from the web, and how this balance affects the quality of the outputs.

\subsection{Differences in Number of Sources} %
\Cref{fig:num_links} shows the number of links per query for different datasets and search engines.
The figure shows \textbf{vast differences in reliance on external knowledge.}
For instance, the median number of links averaged over all datasets for \gptnormal is $0$ as compared to $9$ for \aio.
\gptsearch retrieves the second-lowest number of links. The median number of links averaged over all the datasets is $4$.

The reliance on external links is query-specific.
While the median number of links for \aio is similar to the median number of links for \organic ($9$ \vs $10$), they differ sharply at the extremes. At the $10^{\text{th}}$ percentile, \aio retrieves fewer links ($0.6$ \vs $10$), while at the $90^{\text{th}}$ percentile, it retrieves far more ($17$ \vs $10$). In other words, \organic search is effectively fixed at $10$ results, whereas \aio adaptively decides how many web pages to retrieve.

We analyze queries for which the \aio retrieved more than $30$ web pages (about $2\%$ of all queries). 
These queries tend to be open-ended in nature, \eg ``What jobs will AI agents replace?'' and ``How can I calm down myself''.
On the other hand, queries where \aio retrieved very few, \eg $2$ or fewer web pages ($15\%$ of AIOs) tend to be short and fact-seeking,
\eg ``what is alabama state university motto'', ``Where is the capital of South Korea?'').
The results show that \textit{for static, well-accepted facts, generative search can potentially retrieve the information from the model's internal knowledge, saving the effort} spent in retrieving $10$ \organic search results, and the effort on the users' part to process them. We now explore this potential in more detail.

\subsection{Retrieval Efficiency on Static Facts}
\label{sec:ret_efficiency}

Retrieval is not free---it incurs additional cost and latency.
In this section, we compare the efficiency of various generative engines by studying their retrieval patterns on well-known facts, \ie facts that frontier models are expected to recall from their internal memory alone~\cite{liang2023holistic}.

We use a special-purpose dataset of simple factual queries with static answers.  The dataset consists of 249 questions, each asking for the capital of a country or territory.
Queries follow the template ``What is the capital of XX?''.
Full dataset construction details and prompt templates are provided in Appendix~\ref{app:capitals}.

As shown in \autoref{tab:retrieval-behavior},
all engines except \aio achieve $100\%$ accuracy.
\aio produces a single error, mistaking the capital of the country Georgia for Atlanta, the capital of the US state of the same name.
\gptnormal does not perform a single search yet answers all queries correctly, hinting at the potential for enhancing efficiency for simple queries related to static facts.
The results show that \textbf{generative search engines frequently perform web searches even for queries answerable from internal knowledge alone.}

\begin{table}[t]
    \centering
    \resizebox{\linewidth}{!}{
    \begin{tabular}{lccc}
        Engine & $\%$ queries with search & mean \#URLs & Accuracy\\ \midrule
        \gptnormal  & 0 \%  & 0      & 100\%\\
        \gptsearch  & 98\%  & 5.2    & 100\%\\
        \gemini     & 100\% & 4.47   & 100\%\\
        \sonar      & 100\% & 8.66   & 100\%\\
        \aio        & --     & 5.83   & 99.5\%\\
    \end{tabular}
    }
    \caption{Retrieval behavior on simple factual queries.}
    \label{tab:retrieval-behavior}
\end{table}

\subsection{Efficiency on Time-Sensitive Queries} %
\label{sec:trending}
While some queries can be answered from internal, static knowledge, others depend critically on recent information.
We therefore turn to time-sensitive queries, \ie the \textbf{Trends dataset} (\S\ref{sec:datasets}), to examine how generative search engines behave when queries require recent, up-to-date information.\footnote{We did not run \sonar for the Trends analysis.}

\begin{figure}[t]
    \centering
    \begin{subfigure}[t]{0.9\linewidth}
        \centering
        \includegraphics[width=\linewidth]{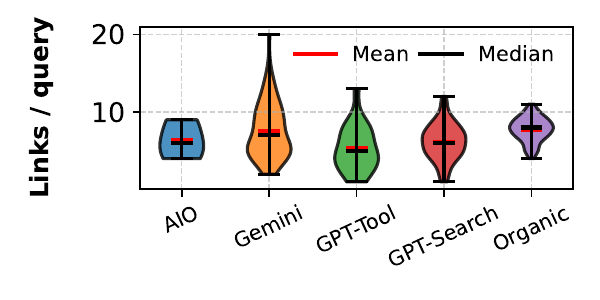}
    \end{subfigure}
    \hspace{\fill} 
    \begin{subfigure}[t]{\linewidth}
        \centering
        \includegraphics[width=\linewidth]{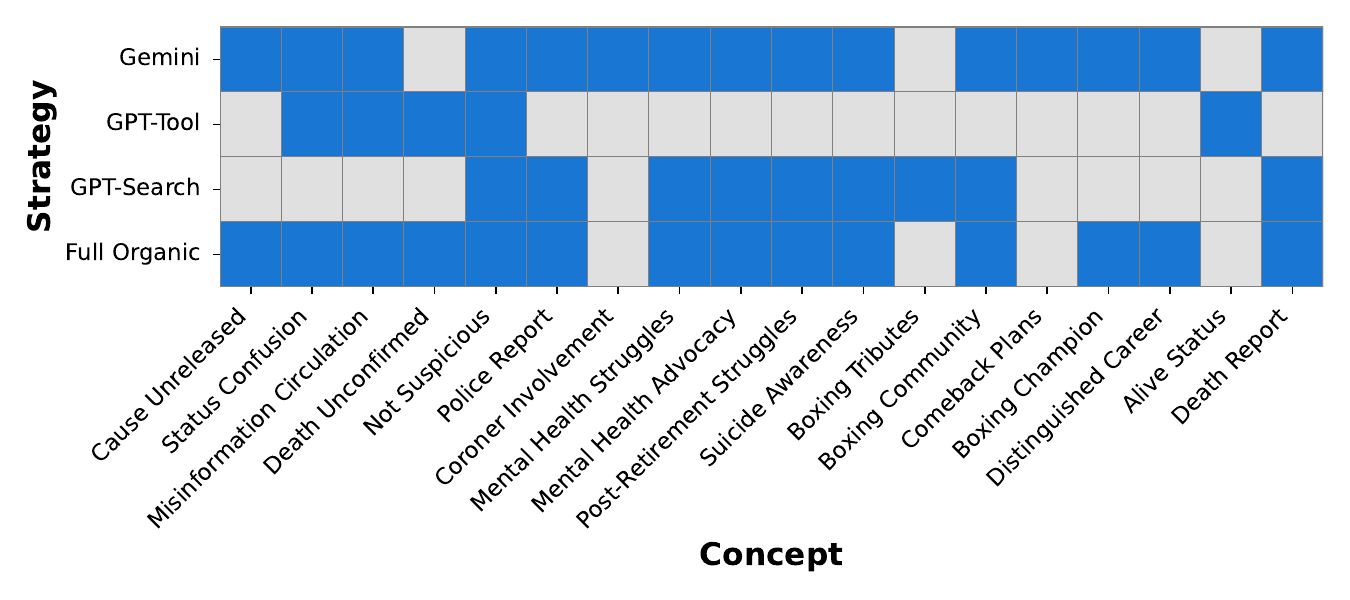}
    \end{subfigure}    
    \caption{\textit{Top:} Number of retrieved links on the trending queries dataset. \textit{Bottom:} Concept coverage of different search engines on the query ``ricky hatton cause of death''. Low concept coverage shows that \gptnormal does not have all the information available.}
    \label{fig:trending}
\end{figure}

We find large structural differences w.r.t. static queries (\S\ref{sec:ret_efficiency}).
\Cref{fig:trending}, top panel, shows the number of retrieved links per query.
The median number of retrieved links is $5$ or more for all engines. \gptnormal still retrieves the least number of links.

We next measure the number of distinct concepts covered by each generative engine.
Methodological details of the concept detection procedure are provided in \S\ref{sec:content_analysis}.
\gptsearch achieves the highest average coverage ($72\%$), followed by \organic ($67\%$) and \gemini ($66\%$), while \gptnormal lags behind at $51\%$.
Manual inspection of queries where \gptnormal exhibits low coverage shows that the model often fails on fast-changing or event-driven topics that require external retrieval.
For example, for the query ``ricky hatton cause of death'', \gptnormal lacks access to recent information, becomes uncertain, and incorrectly reports Ricky Hatton as alive (see \Cref{fig:trending}, bottom panel).
To assess robustness, we collected additional trend snapshots on two other days. The relative patterns remain consistent across snapshots.

Findings of \S\ref{sec:ret_efficiency} and \S\ref{sec:trending} highlight a \textbf{tension between efficiency and accuracy} in generative search.
For simple, static factual queries, the optimal response may simply be the correct entity name.
Long answers or extensive web searches may not add value.
However, over-reliance on internal knowledge can lead to outdated answers when information changes over time.
Correctly differentiating between static facts and dynamic knowledge can help mitigate this tradeoff.

\section{Source Breadth \& Content Diversity}
\label{sec:knowledge_breadth}
Reliance on internal \vs external knowledge alone does not determine what information users ultimately see. 
We now examine the popularity, ranking depth, and diversity of cited sources to understand how generative search engines reshape the informational landscape relative to traditional search.

\subsection{Source Popularity and Rank Depth}

\begin{figure}[t]
        \centering
        \includegraphics[width=0.95\linewidth]{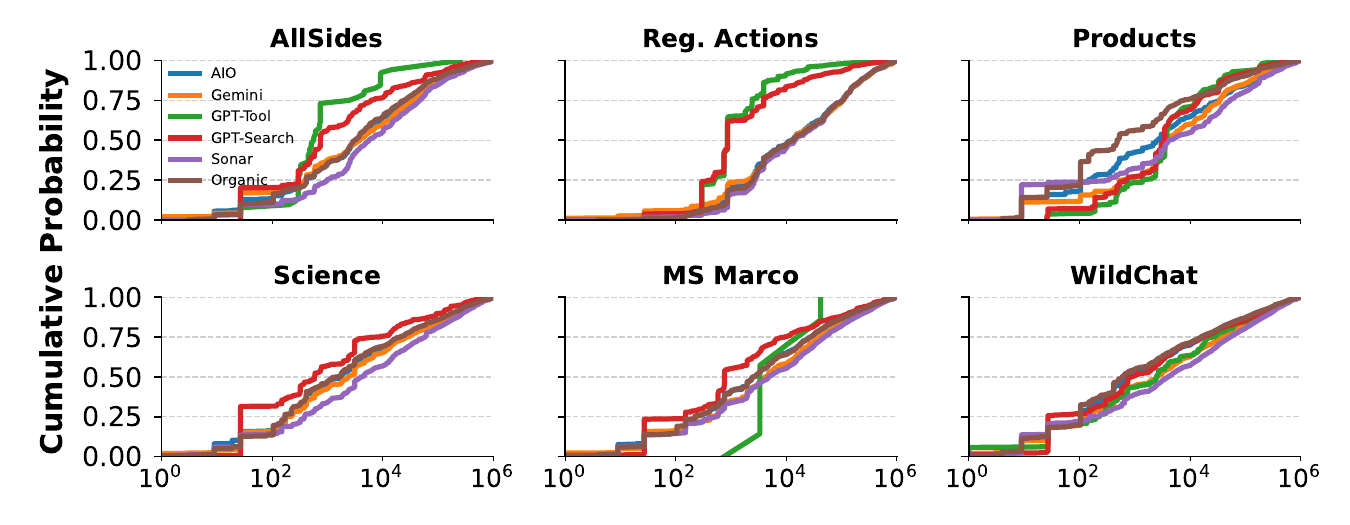}
    \caption{
    Different search engines select information from domains of differing ranks.
    }
    \label{fig:link_cdf_us}
\end{figure}

\xhdr{Most generative engines cite sources of lower popularity than \organic search.}
We use the Tranco rankings%
\footnote{\url{https://tranco-list.eu} (Retrieved in July 2025)}
to measure the rank of sources within the 1M most visited domains.
$89\%$ of \organic sites appear in the top-1M list, compared to $85\%$ for \aio, $86\%$ for \gptsearch, $84\%$ for \sonar, $83\%$ for \gemini, and $81\%$ for \gptnormal.
\Cref{fig:link_cdf_us} shows the rank CDF for the links that were found within the Tranco 1M list.
While \gptnormal retrieves fewer popular sites overall, the sources it does cite are often highly ranked: its median domain rank ($1{,}124$) exceeds that of \organic ($2{,}352$), except on the Products dataset.
For the \regactions dataset, both the GPT models retrieve sources that are ranked significantly higher. 
\sonar cites the least popular sources overall, with a median domain rank of $5647$.

\begin{figure}[ht]
    \centering
    \includegraphics[width=.9\linewidth]{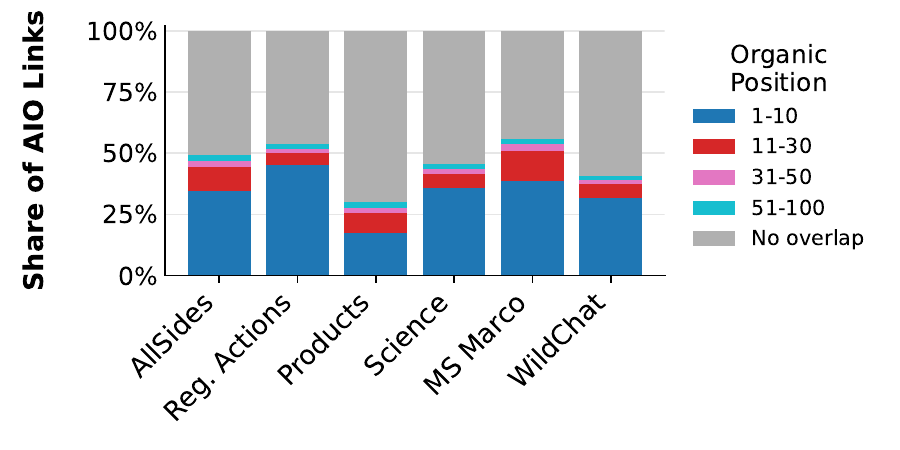}
    \caption{Overlap between retrieved links and top-100 \organic links for \aio. More than $40\%$ of \aio's links are not contained in the top-100 \organic links.}
    \label{fig:reach_100_stacked_aio}
\end{figure}

\xhdr{Generative search reaches far beyond \organic search.}
\Cref{fig:reach_100_stacked_aio} shows that the \aio links have less than $50\%$ overlap with the top-10 \organic results, and overlap remains below $60\%$ even when considering the top-100.
\Cref{fig:reach_100_stacked_gpts_gemini_us}
in \Cref{app:link_analysis} shows the overlap for other search engines. 
Overlap is substantially lower for \gemini and \gptsearch, and close to zero for \gptnormal.
Domain-level overlap is higher than URL-level overlap, but a substantial fraction of cited domains still fall outside the top-100 \organic domains (Figure~\ref{fig:reach_100_stacked_domain_gpts_gemini_us}).
These findings indicate that generative search significantly broadens the set of consulted sources.

\subsection{Source Diversity}

\xhdr{Generative engines draw from different types of sources than \organic search.}
We classify domains into high-level categories using the Google Content Categories (26 categories) and a more web-oriented hand-designed categorization (10 categories) using an LLM as a judge. Details can be found in \Cref{app:category_classification}.
The hand-designed list includes categories like ``Encyclopedia'' (\eg Wikipedia and World Atlas) and ``Science Publisher'' (\eg to classify websites like ACM, ScienceDirect and arXiv).

\Cref{fig:links_categories_us_custom,fig:links_categories_us_google} in \Cref{app:knowledge_breadth}  show large differences in category distribution of websites retrieved by various engines.
The GPT models rely heavily on Corporate Entities and Encyclopedias, with less reliance on Social Media and User Forums.
In contrast, other engines like \organic can retrieve up to $35\%$ from such websites.
In summary, not only does it expand the depth of retrieval, \textbf{generative search also reshapes the composition of sources} to which users are exposed.

\begin{figure}[t]
    \centering
    \includegraphics[width=.95\linewidth]{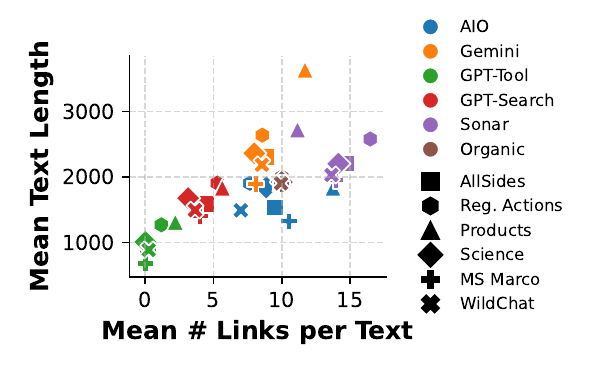}
    \caption{
    Ratio between the retrieved number of links and text length.
    \gptnormal returns the shortest texts, but also retrieves the fewest URLs.
    }
    \label{fig:len_vs_link}
\end{figure}

\subsection{Conceptual Diversity} %
\label{sec:content_analysis}

We now study how differences in sources translate into differences in the content presented to users.

\subsubsection{Output Characteristics}
\xhdr{Generative engine outputs differ structurally from \organic search and between engines.}
Across generative engines, output length and the number of links vary widely.
For \gptnormal, \gptsearch, and \gemini, longer responses tend to include more links, whereas \aio often produces comparatively short texts with many citations (\Cref{fig:len_vs_link}).
\gemini produces the longest responses on average ($2284 \pm 1239$ characters), with especially long outputs for \prods (mean: $3636 \pm 995$), while \gptnormal generates the shortest texts (mean: $939 \pm 575$) and cites the fewest links ($0.25$ per search result).
We next examine how these differences shape the topical breadth of the content.

\subsubsection{Topical Content Analysis}

We compare the high-level concepts mentioned per search engine for each query.
We use \lloom~\cite{Lam_2024}, an LLM-powered topic inference framework that annotates unstructured text with interpretable concepts (details in  \Cref{app:lloom}).

\begin{figure}[t]
    \centering
    \includegraphics[width=.98\linewidth]{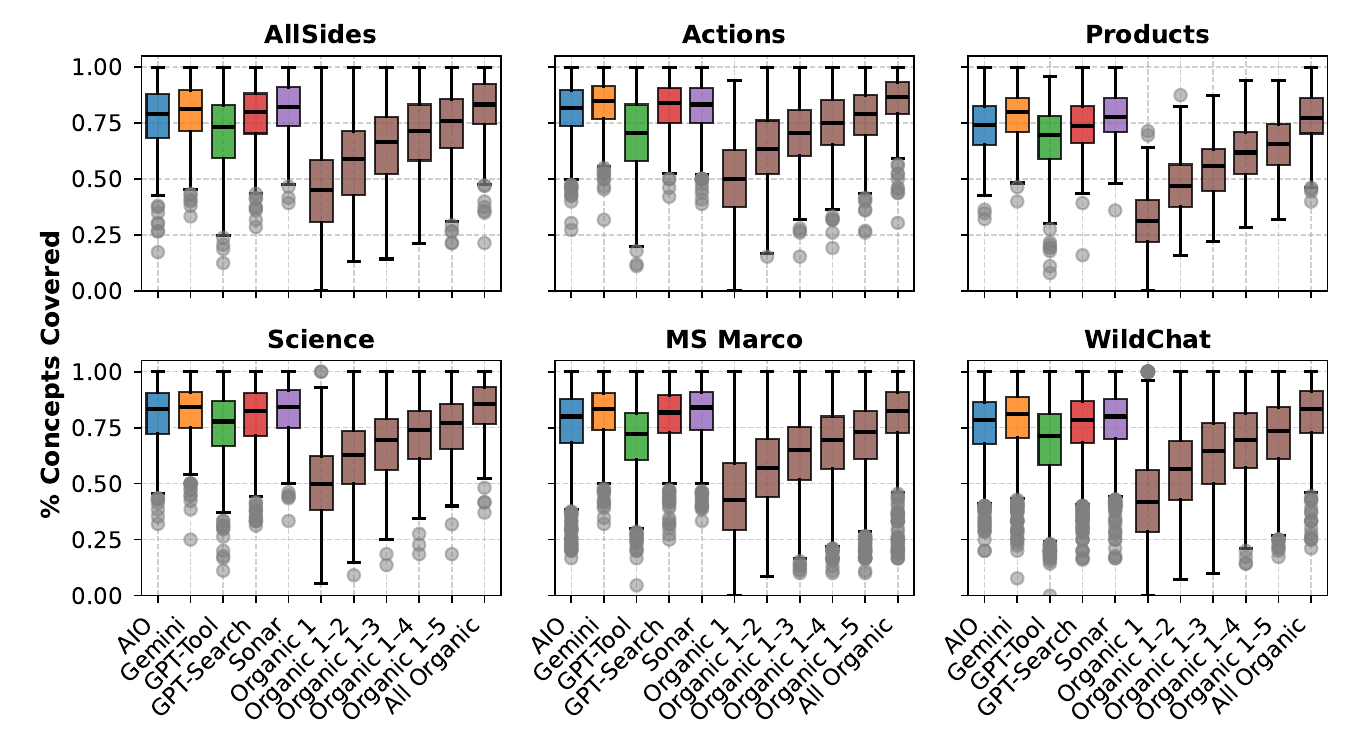}
    \caption{Fraction of concepts covered per engine. \organic reaches high coverage within five results.}
    \label{fig:concepts_per_strategy}
\end{figure}

\xhdr{Traditional and generative search achieve similar overall topic coverage.}
Across datasets, generative engines achieve coverage comparable to \organic search (\Cref{fig:concepts_per_strategy}).
For instance, average coverage ranges from $0.71$ (\gptnormal) to $0.78$ (\gemini), closely matching \organic ($0.78$).
In most cases, the first five organic search results are sufficient to achieve high coverage, with cumulative gain showing diminishing returns. 
This observation raises a key design question: {\it When generative search can condense objective information into concise responses, how many organic snippets does a user actually need to see?}

While aggregate coverage values are similar, the specific concepts mentioned differ across engines.
The average concept overlap between generative engines and \organic ranges only from $60\%$ to $68\%$ per query (\Cref{fig:overlap_cdf_lloom_us} in \Cref{app:lloom}).
At the $10^\text{th}$ percentile of queries, where agreement across engines is lowest, \organic reaches a median coverage of $67\%$, compared to $55\%$ for \aio and $48\%$ for \gptnormal.
Queries in this low-agreement regime tend to be underspecified (\eg ``when did Queen Elizabeth''), malformed (\eg ``when is a gene''), or ambiguous (\eg ``what is an example of inequality?'').
Figure~\ref{fig:heatmap_query_0} in \Cref{app:lloom} illustrates an example in which different engines surface largely non-overlapping concepts for the same query.
In other words, \textbf{Organic search retains an advantage on ambiguous queries.}

Overall, this analysis suggests that, despite accessing more, and often lower-ranked, sources, generative search engines do not consistently surface more concepts than traditional search.

\section{Stability of Generative Search}
\label{sec:stability}
Generative search systems introduce new challenges around the \textit{stability} of search outputs.
They combine evolving web content, complex retrieval pipelines, and stochastic language models.
As a result, identical queries may yield different outputs depending on \textit{when} and \textit{how} they are executed.
We study two complementary sources of variability:
\textbf{temporal drift}, arising from changes in models and web content over time, and \textbf{stochasticity}, which can induce variation even across closely spaced executions of the same query.

\subsection{Temporal Stability} %
\label{sec:temporal_stability}
To quantify how outputs evolve as the web and models update, we repeated all experiments approximately two months apart (July/August and September 2025).

The number of triggered AIOs and conducted searches for \gptnormal shows minor changes (Table~\ref{table:data_stats} in Appendix~\ref{sec:data_stats}).
We also do not observe any notable changes in the number of links retrieved by different engines, with the only notable exception being \aio where the average number of links per query changes from $8.4$ to $7.2$.

Rank distributions remain mostly consistent for \organic search, but vary considerably for generative systems, especially \gptnormal and \gptsearch, whose retrieved domain popularity shifts by as much as 40,000 ranks (Figure~\ref{fig:newold_ranks} in Appendix~\ref{app:newold}).
Comparing the Jaccard similarity between sets of retrieved links across runs (\Cref{fig:newold_overlap} in Appendix~\ref{app:newold})
shows highest overlap for \organic search ($45\%$). 
In contrast, \aio exhibits substantially lower overlap ($18\%$) between runs. 
This observation aligns with prior reports indicating that Google’s AI mode often returns markedly different results across repeated sessions.\footnote{\url{https://tinyurl.com/aio-differences}}
Despite this source-level variability, overall conceptual coverage remains largely stable across time (\Cref{fig:newold_concepts} in Appendix~\ref{app:newold}), indicating that generative systems can maintain a similar topic coverage in their generated summaries even as retrieved sources change.

\subsection{LLM Stochasticity} %
\label{sec:stochasticity}

Generative search engines rely on stochastic language models and dynamic retrieval pipelines.
To assess whether this stochasticity materially affects search outcomes, we curate a custom dataset from WildChat~\cite{zhao2024wildchat1mchatgptinteraction} that allows only three possible answers: affirmative, negative, or mixed. The dataset consists of questions like ``Is it rare for K-pop songs nowadays to have full Korean track titles?''. We provide details on dataset construction in \Cref{app:yes_no_data}.

Our goal is to study whether the stochasticity of the underlying LLM changes the polarity of the response, \eg from affirmative to negative.
To that end, we issued the queries at three time points ($t_0$, $t_{5m}:=t_0 + 5$ minutes, and $t_{24h}:=t_0 + 24$ hours).
We study two temperature conditions: (i) a zero-temperature setting, and (ii) the default temperature per model ($1.0$ for \gptnormal and \gemini, $0.2$ for \sonar, temperature control is not supported for \gptsearch and \aio). %
Since generative models can provide an affirmative response without mentioning the term ``yes'', we use an LLM-based judge to classify each response as \textit{affirmative}, \textit{negative}, or \textit{mixed} (details in Appendix~\ref{app:yes_no_judge}).
Table \ref{tab:example_temporal_variation} shows example responses with judge annotations.

\begin{table}[t]
\centering
\small
\resizebox{\linewidth}{!}{
\begin{tabular}{l p{6cm} l}
\toprule
& \textbf{Response} & \textbf{Label}\\
\midrule
$t_0$ 
& \textbf{No,} the Province of Soria does not directly border Navarre; Soria is in central Spain, bordered by La Rioja, Zaragoza, Guadalajara, Segovia, and Burgos, ...
& (\textit{negative}) \\

$t_{5m}$ 
& \textbf{Yes,} the Province of Soria (in Castile and León) and the region of Navarre share a border, with Soria touching Navarre and Aragon to its east/northeast, ...
& (\textit{positive}) \\

$t_{24h}$ 
& \textbf{Yes,} the Province of Soria in Spain does border Navarre; they share a northern boundary, ...
& (\textit{positive}) \\
\bottomrule
\end{tabular}
}
\caption{
\aio responses for the query ``Does the Soria province have a border with Navarre?''.
Both wording and final answer polarity change over time.
}
\label{tab:example_temporal_variation}
\end{table}

\begin{table}[t]
\centering
\resizebox{\linewidth}{!}{
\begin{tabular}{lcccc}
\toprule
\textbf{Engine} 
& \multicolumn{2}{c}{\textbf{Temp. = 0}} 
& \multicolumn{2}{c}{\textbf{Default Temp.}} \\
\cmidrule(lr){2-3} \cmidrule(lr){4-5}
& $t_0 \rightarrow t_5$ 
& $t_0 \rightarrow t_{24}$
& $t_0 \rightarrow t_5$ 
& $t_0 \rightarrow t_{24}$ \\
\midrule
\gptnormal      & $9\%$   & $10\%$  & $18\%$  & $18\%$ \\
\gptsearch      & $16\%$  & $17\%$  & $15\%$  & $22\%$ \\
\gemini         & $15\%$  & $15\%$  & $20\%$  & $20\%$ \\
\aio            & $17\%$  & $16\%$  & $11\%$  & $17\%$ \\
\sonar          & $27\%$  & $28\%$  & $27\%$  & $24\%$ \\
\bottomrule
\end{tabular}
}
\caption{
Percentage of queries with a flip in repeated executions (between \textit{affirmative}, \textit{negative}, \textit{mixed}).
Results are shown separately for zero temperature and default temperature settings,
and for short-term ($t_0 \rightarrow t_5$) and longer-term ($t_0 \rightarrow t_{24}$) time gaps.
}
\label{tab:decision_flips}
\end{table}

\xhdr{Decision Stability.}
Table \ref{tab:decision_flips} reports the fraction of queries for which the overall decision changes.
We observe \textbf{substantial decision instability in decisions} across all generative search engines.
Even at  $T=0$, between $9\%$ and $27\%$ of queries exhibit a decision flip within five minutes,  with slightly higher rates observed over 24 hours.
Lexical overlap across repeated executions is low for all engines.
Mean Jaccard similarity ranges from  $0.27$ to $0.63$ at zero temperature (Table~\ref{tab:jaccard_similarity} in Appendix~\ref{app:textual_overlap}), indicating that repeated executions often share less than half of their lexical content.
Increasing the temperature further amplifies textual variability for most engines that have temperature control.

\xhdr{Comparison to Organic Search.}
We compare the answer polarity of \organic search with generative search. 
While generative search output consists of one coherent piece of text, \organic search shows around 10 results on the first page. 
We label each of the top-10 search results individually.
Across all generative engines, $55\%$ of the responses are labeled as affirmative, $19\%$ as negative, and $26\%$ as mixed.
In contrast, in only $16\%$ of queries, all top-10 organic search results express a single polarity label (affirmative, negative, or mixed).
These results highlight a structural difference between search paradigms:
\textit{
Organic search distributes ambiguity and disagreement across multiple results.
Generative search resolves this plurality into a single answer that may itself change over time.
}

\section{Conclusion and Future Work}
We conduct a systematic evaluation of generative search engines, focusing on new dimensions that are largely absent from traditional ranked-list evaluation.
By comparing a conventional search engine with multiple generative search engines across diverse datasets, we show that generative search introduces trade-offs in how information is retrieved, synthesized, and presented.
In particular, systems vary substantially in their reliance on internal \vs external knowledge, retrieval efficiency, knowledge breadth, and stability. 
Despite often achieving comparable topical coverage, generative engines differ markedly in their retrieval footprints, source diversity, and sensitivity to time and stochasticity.

These findings suggest that evaluating generative search requires moving beyond correctness-centric metrics towards benchmarks that explicitly account for sourcing behavior, synthesis, and variability over time.
At the same time, traditional search and generative search each exhibit complementary strengths:
While ranked lists of results can expose multiple perspectives and offer greater stability, generative search enables aggregation from a wider range of sources and, in some cases, more efficient answers by leveraging internal knowledge.
Understanding and measuring these trade-offs is critical as generative search systems increasingly serve as interfaces to online information.

This work opens several directions for future work.
User-centric evaluations could study how different search paradigms affect trust, satisfaction, and decision-making.
The observed temporal and stochastic variability motivate longitudinal evaluations that track system behavior over time and across model updates.
Finally, evaluating a broader range of models and retrieval architectures could shed light on how different design choices shape retrieval strategies and synthesis behavior.

\newpage
\section{Limitations}
Our work has several limitations.
First, the scope of our analysis is restricted to selected query workloads spanning general information, products, politics, and science.
We did not consider multi-turn conversational searches, and all experiments used English-language queries in only two countries.

Second, our analysis of \organic search content is limited to the first ten results, assuming users rarely go beyond the first page. While prior work provides evidence to support this assumption~\cite{10.5555/1484611.1484615,stacy_search,urmanYouAreHow2023}, it may not hold for niche cases.
We also restrict our analysis to the title, URL, and snippet of each search result, assuming that users rarely click on links.
While there is some evidence supporting this behavior~\cite{miroyan2025searcharenaanalyzingsearchaugmented}, we acknowledge that this design omits the information contained in the full underlying webpages and ignores possible user actions that deviate from these assumptions.
However, as shown in our extended analysis with full crawled webpages in Appendix~\ref{app:supp_fullweb}, this approximation primarily affects a recall–precision trade-off: full webpages increase concept recall but also introduce additional, often less query-relevant content.

Third, our evaluation captures a limited time window, while both organic and generative outputs evolve over time with model updates, indexing changes, and emerging events.
In addition, while we controlled for several experimental factors (\eg logged-out sessions, standardized query execution, and fixed geographic settings), we cannot fully eliminate all sources of variability inherent to commercial search systems, such as backend personalization, infrastructure differences, or time-of-day effects.
These factors may introduce additional noise into the observed outputs.
Furthermore, we examine a limited set of search engines.
Our study focuses on deployed, closed generative search systems. 
Moreover, generative outputs are inherently non-deterministic, and our evaluation uses only one output per query, though we used a temperature value of $0$ whenever configurable.

Finally, our content analysis relies on an LLM-based concept induction method (\lloom), which may introduce potential bias and can reflect the model's own knowledge gaps or preferences.
While human annotation could verify these concepts, it does not scale well to large datasets.
Future work should explore scalable alternatives, such as semi-supervised methods (\eg labeled LDA), and combine coverage metrics with fact-checking and credibility assessments.

\section{Ethical Considerations}
We believe this study does not raise direct ethical concerns or potential risks.
The \wildchat dataset we used contains real-world user-LLM interactions. However, any sensitive or harmful content was removed by the dataset authors prior to our use.
All other queries in our evaluation were drawn from publicly available datasets, commercially available product lists, or were generated synthetically for the purpose of controlled experiments.
Our analysis focuses exclusively on publicly accessible search results and model outputs, without altering or influencing any live search systems.

\section{Acknowledgments}
We thank Nicole Krämer for her helpful feedback.

\xhdr{Use of Generative AI.}
Asta\footnote{\url{https://asta.allen.ai/}} was used for literature search. However, all used references were reviewed, summarized, and added manually.
Cursor\footnote{\url{https://cursor.com/}} was used for coding assistance.
LLMs were not used to directly generate any manuscript text. 
We used LLMs (GPT-5.2) for rephrasing and shortening select statements.

\bibliography{main}
\balance
\clearpage
\newpage
\appendix

\nobalance

\section{Datasets}
\label{app:datasets}

\subsection{Additional Details on Dataset Construction}
Our goal was to select datasets that cover one or more of the following desiderata. The datasets should:
(i) reflect real users' queries to both traditional and generative search engines,
(ii) cover everyday queries as well as more domain-specific workloads like politics, science, and shopping, and
(iii) not only focus on queries relating to persistent topics (\eg ``nemesis in literature''), but also focus on time-dependent topics, \eg  ``Designating English as the official language of the US'' which was an executive order issued in March 2025.

\xhdr{\regactions.} 
For each executive order, we ask GPT-4o to generate 10 questions that a real person might ask about the order.
The prompt for generating the queries about the executive orders (\Cref{sec:datasets}):

\begin{tcolorbox}[
    enhanced, 
    breakable,
    fontupper=\footnotesize\ttfamily,
    left=1pt,
    right=1pt,
    top=1pt,
    bottom=1pt,
    standard jigsaw,
    opacityback=0
]
You are a helpful assistant that generates realistic questions that people might ask about regulatory actions.
\\
\\
Please generate \{num questions\} different questions that a real person might ask about this regulatory action: \{action\}
\\
\\
The questions should be:\\
- Natural and conversational (like how someone would actually ask)\\
- Focused on practical concerns, implications, or clarifications that a person might ask about\\
- Short and direct (less than 15 words)\\
- Self-contained (include the name of the action in each question)\\
- Different from each other
\\
\\
Format your response as a JSON array of strings, like this:
[``Question 1'', ``Question 2'', ``Question 3'', ...]
\\
\\
Respond only with the JSON array, no additional text.
\end{tcolorbox}

\begin{table*}[ht]
    \centering
    \resizebox{\linewidth}{!}{
    \begin{tabular}{l cc  cccc c cccc}
        \toprule
        & & & \multicolumn{4}{c}{\textbf{September 2025}} && \multicolumn{4}{c}{\textbf{July 2025}} \\
        \cmidrule{4-12}
        & & & \multicolumn{2}{c}{\textbf{\#AI Overviews}} & \multicolumn{2}{c}{\textbf{\# Searches GPT}}
        && \multicolumn{2}{c}{\textbf{\#AI Overviews}} & \multicolumn{2}{c}{\textbf{\# Searches GPT}} \\
        \cmidrule{4-7} \cmidrule{9-12}
        {\bf Dataset}
        & {\bf \#Items}  
        & {\bf \#Queries} 
        & {\bf US} 
        & {\bf DE} 
        & {\bf US}
        & {\bf DE}
        &
        & {\bf US} 
        & {\bf DE} 
        & {\bf US}
        & {\bf DE}
        \\
        \toprule
        {\bf \wildchat}    & 1M      & 1,750 & 1,425 & 1,050 & 97  & 63  && 1,007 & 1,031 & 73  & 66  \\
        {\bf \msmarco}     & 100K    & 1,000 & 847   & 784   & 4   & 2   && 792   & 780   & 3   & 3   \\
        {\bf \prods}       & 100     & 422   & 160   & 42    & 64  & 5   && 124   & 46    & 59  & 1   \\
        {\bf \allsides}    & 37      & 332   & 280   & 233   & 10  & 6   && 241   & 235   & 8   & 11  \\
        {\bf Reg. Actions} & 58      & 649   & 596   & 437   & 136 & 100 && 427   & 447   & 110 & 104 \\
        {\bf Science}      & 45      & 453   & 418   & 315   & 1   & 0   && 398   & 361   & 1   & 1   \\
        {\bf Trends}      & 100      & 100   & 3   & 1   & 19   & 19   && --   & --   & --   & --   \\
        \bottomrule
    \end{tabular}
    }
    \caption{
    Statistics for each dataset. Experiments were conducted in July/August and September 2025. \#Items shows the number of items in the source dataset. \#Queries shows the number of queries we obtained after the processing steps performed in \Cref{sec:datasets}. \#AI Overviews shows the number of Google AI Overviews generated for US and Germany (DE).
    \# Searches GPT shows the percentage of times \gptnormal performed a web search.
    }
    \label{table:data_stats}
\end{table*}

\xhdr{\prods.}
We turn the list of gathered products into review and comparison-oriented questions using custom templates like ``<product name> review'' and ``<product name> worth it''.

Given an item from the top-100 Amazon search terms, we use a GPT-4 model to annotate each product with categories and subcategories from Amazon's own category system (e.g., Electronics, Computers, Arts \& Crafts), and to generate potential use cases which represent real-world needs or scenarios for the item.
We manually extracted the brand names from items that mention a brand.

Given a search term, the usecase, the corresponding category, subcategory and the brand name, we manufacture the search queries as follows:
\begin{enumerate}
    \item <product name> reviews
    \item <product name> worth it?
    \item alternatives to <product name>
    \item best <product name>
    \item best <subcategory name> for <use case>
    \item best <use case> <subcategory name>

\end{enumerate}

\subsection{Dataset-level Statistics}
\label{sec:data_stats}
\Cref{table:data_stats} reports the total number of data points, number of queries that we randomly selected for search,  number of queries that lead to AI Overviews and the percentages of these that triggered a search in \gptnormal.

\section{Additional Information on Experimental Setup}
\label{app:experimental}

Our desiderata for selecting search engines are to:
(i) study both traditional and generative search engines,
(ii) cover generative models whose dedicated use is to be a web search engine {\it as well as} generative models that are mainly chatbots but can use web search as a tool to respond to user queries. The latter is motivated by the fact that users are increasingly using chatbots to perform tasks similar to web search~\cite{HYUNBAEK2023102030,sanders_hbr}.
We focus on deployed, user-facing search systems to characterize how generative search behaves in practice. Our objective is not to evaluate models in isolation, but end-to-end systems that integrate retrieval, ranking, and synthesis. Because most open-source models still lack comparable, interchangeable search pipelines, directly comparing them to closed systems would conflate model capability with system design.

We use the following search engines:

\xhdr{Organic Google Search (\organic).}
We query the Google search engine to obtain the search results, setting the number of top results to be retrieved to $100$. 
In order to compare the overlap between generative search and Google search at various ranks, we analyze these top-100 returned results the \organic search.
For the remaining analysis, we only consider the top-10 results.
In rare cases, Google returns fewer than 10 results due to the presence of other content, like Google Knowledge Graph results. We use the SERP API to automatically query Google websites.%
\footnote{\url{https://serpapi.com/search-api}}
We controlled fixed IP regions by setting up virtual machines in the respective countries (US, DE). Both \organic and \aio were executed using the same SERP API requests, ensuring identical request parameters. All queries were executed independently and in parallel in logged-out sessions using clean virtual environments.

\xhdr{Google AI Overviews (\aio).}
In addition to the organic search results, a Google search in most cases also results in an AI Overview. Whether to generate an AI Overview is decided internally by Google Search.
The AI Overviews are also accompanied by source URLs that were used to generate the AI overview.
Overall, $81\%$ of the queries generated an AI overview in the US location and $65\%$ of the queries generated an AI overview for the DE location.
See Table~\ref{table:data_stats} in \Cref{app:datasets} for the number of cases per dataset.
\textit{We perform the search comparisons for queries to which all the search engines (organic and generative) produce an answer.}
Hence, we query the remaining models only with inputs where \aio generated a response, ensuring interface-level comparability.
To assess generalizability, we conducted additional analyses on the full query set and observed qualitatively consistent trends.
We observe only small relative differences between the full set and the AIO-conditioned subset in terms of retrieval footprints, with changes in the mean number of links typically within $\pm$10\%, with differences falling within overlapping $95\%$ confidence intervals. The average difference in median rank was up to $527$ (for \gptnormal).
Similarly, when re-running topic analysis on the full query set on the \allsides dataset, differences remain minor (maximum change of $+4\%$ for \organic), suggesting that our conclusions generalize beyond the conditioned subset.

\xhdr{Gemini-2.5-Flash with Google Search (\gemini).}
While Gemini-2.5-Flash is primarily an AI assistant, it can be used with web search to ground the results~\cite{gemini_ground}.
Given a query, the model first decides if a web search should be performed.
For our datasets, the model performs the search in more than $99.5\%$ of the cases.
If the model decided to perform a search, it transforms the original query by rephrasing it one or more times. For instance, the query ``What are the main goals of the Make America Healthy Again Commission?'' is transformed to search queries ``Make America Healthy Again Commission goals'' and ``Make America Healthy Again Commission objectives''.
We use a thinking budget of $0$ tokens.

\xhdr{GPT-4o Search (\gptsearch)}
always performs a web search before returning an answer.
We set the search radius to ``medium''.
We used the model version \texttt{gpt-4o-search-preview-2025-03-11}.

\xhdr{GPT-4o with Search Tool (\gptnormal)}
determines on a per-query basis whether to use web search as a tool.
Except for the \prods and \regactions datasets, the percentage of cases when the model performs a search is close to $0$.
See Table~\ref{table:data_stats} in \Cref{app:datasets} for details.
We set the search radius to ``medium''.
Results for different search context size settings are reported in \Cref{app:supp_gensearch}.
The model version we use is \texttt{gpt-4o-2024-08-06}.

\xhdr{Sonar (\sonar)}
always performs a web search when generating an answer. We use the model version \texttt{sonar} with search mode ``web''.

\xhdr{Miscellaneous Parameters.}
All queries were performed in English.
Queries for the generative AI models (\gemini, \gptsearch, \gptnormal, and \sonar) were performed using their respective APIs.
For models that support a temperature parameter (\gemini, \sonar, and \gptnormal), we set it to $0$. 
We set the maximum new tokens to $1,000$.
All queries were performed in September 2025.
To study the effect of location, we performed queries from two locations: United States ({\bf US}) and Germany ({\bf DE}).
We used virtual machines on Google Compute Platform to ensure that the requesting IPs were located in the corresponding country.
The main paper contains the results for the US location, we provide a discussion of the results for the DE location in Appendix \ref{app:de_analysis}.

\section{Supplementary Material for \Cref{sec:internal_vs_external}}
\label{app:capitals}

\subsection{Details on Capitals Dataset Construction}
We construct a dataset of factual queries asking for the capital of countries to study retrieval behavior on simple information needs.
We start from a comprehensive list of 249 countries.\footnote{\url{https://github.com/umpirsky/country-list/blob/master/data/en_US/country.csv}}
For each country, we generate a natural-language query of the form ``What is the capital of XX?''.
To reflect natural English phrasing, we add definite articles where appropriate (\eg ``the Netherlands''), generated using \texttt{gpt-5.2-chat}.
The resulting dataset consists of 249 short queries.
We use the Google Knowledge Graph as the ground-truth source and manually resolve ambiguous cases (\eg countries with multiple administrative capitals).

The prompt used for optionally adding definite articles to the countries using \texttt{gpt-5.2-chat}:
\begin{tcolorbox}[
    fontupper=\footnotesize\ttfamily,
    left=1pt,
    right=1pt,
    top=1pt,
    bottom=1pt,
    standard jigsaw,
    opacityback=0
]
Given the following CSV dataset, create an additional column "question\char`_country" derived from the "country" column.
\\
The "question\char`_country" value must be grammatically correct in the sentence: "What is the capital of \{question\char`_country\}?"
\\
\\
Use the country name as-is when no article is required.  Articles are added only when required in standard English usage, e.g., for plural country names, country names containing "Islands", or political entities commonly used with a definite article (e.g., the United States). Do not change the country names themselves.
\\
\\
Output the complete dataset as a CSV file including the new column.
\end{tcolorbox}

\section{Supplementary Material for \Cref{sec:stability}}
\label{app:stability}

\subsection{Detailed Results from Section~\ref{sec:temporal_stability}}
\label{app:newold}

\begin{figure*}[t]
    \centering
    \begin{subfigure}[t]{0.4\linewidth}
        \centering
        \includegraphics[width=.95\linewidth]{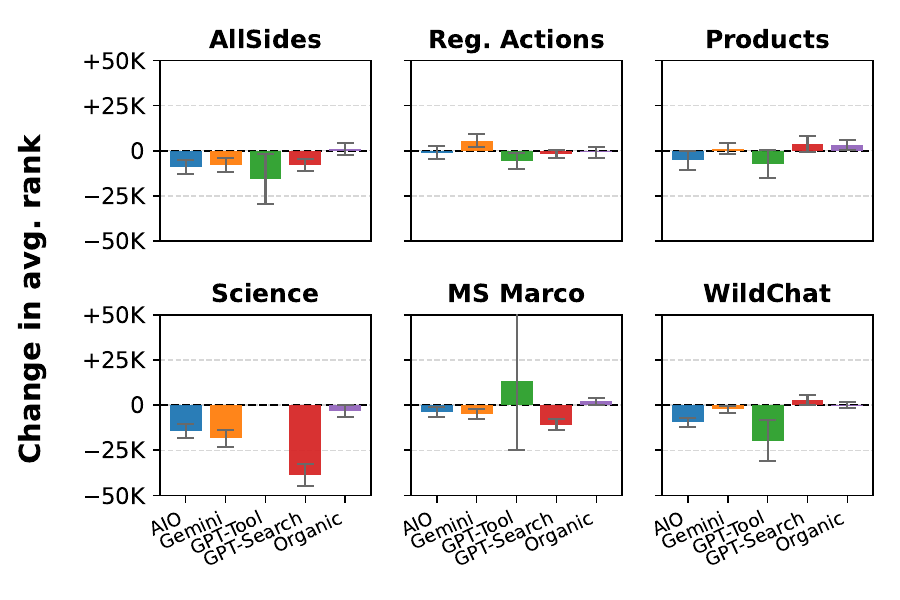}
        \caption{}
        \label{fig:newold_ranks}
    \end{subfigure}
    \begin{subfigure}[t]{0.4\linewidth}
        \centering
        \includegraphics[width=.95\linewidth]{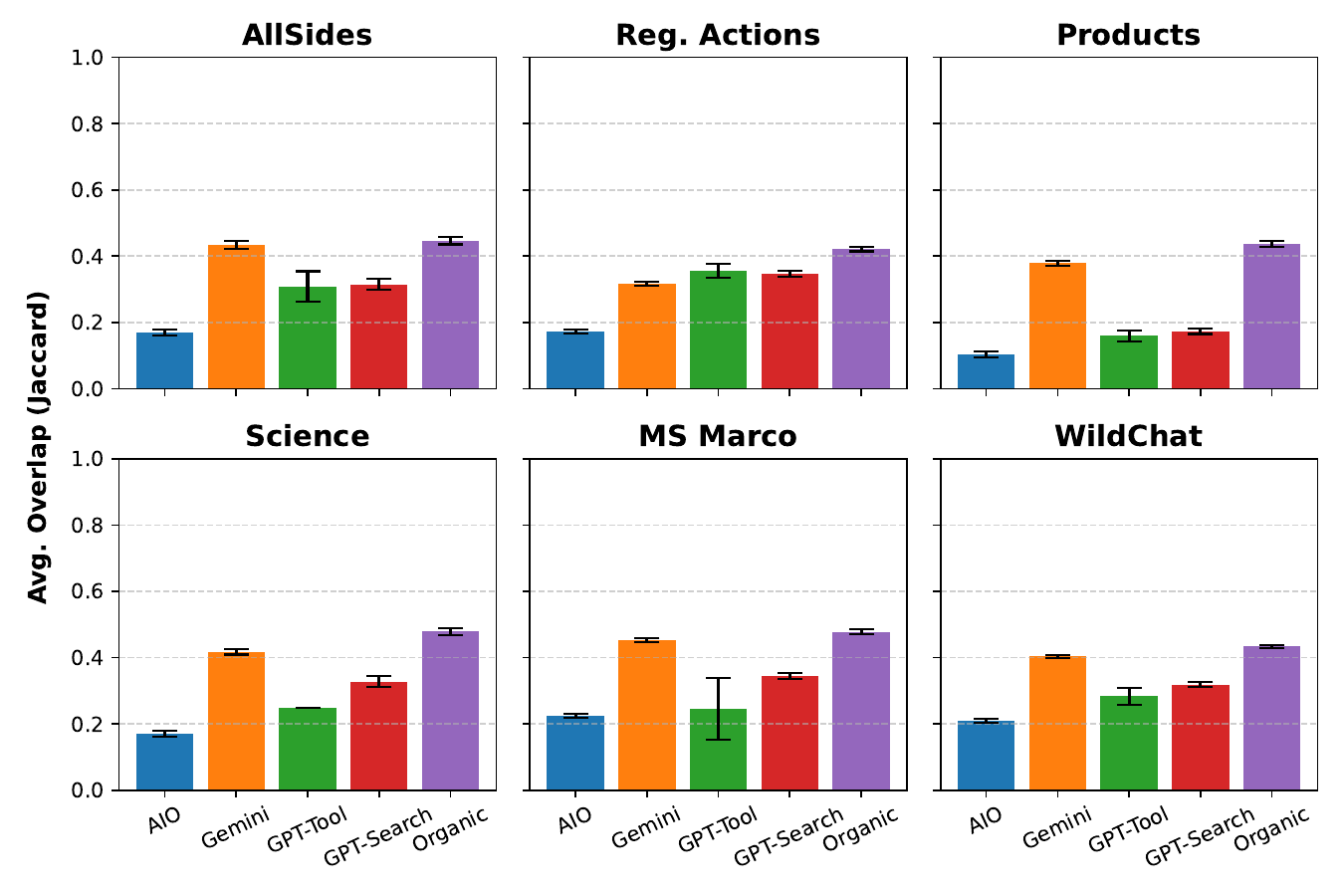}
        \caption{}    
        \label{fig:newold_overlap}
    \end{subfigure}
    \begin{subfigure}[t]{0.7\linewidth}
        \centering
        \includegraphics[width=.95\linewidth]{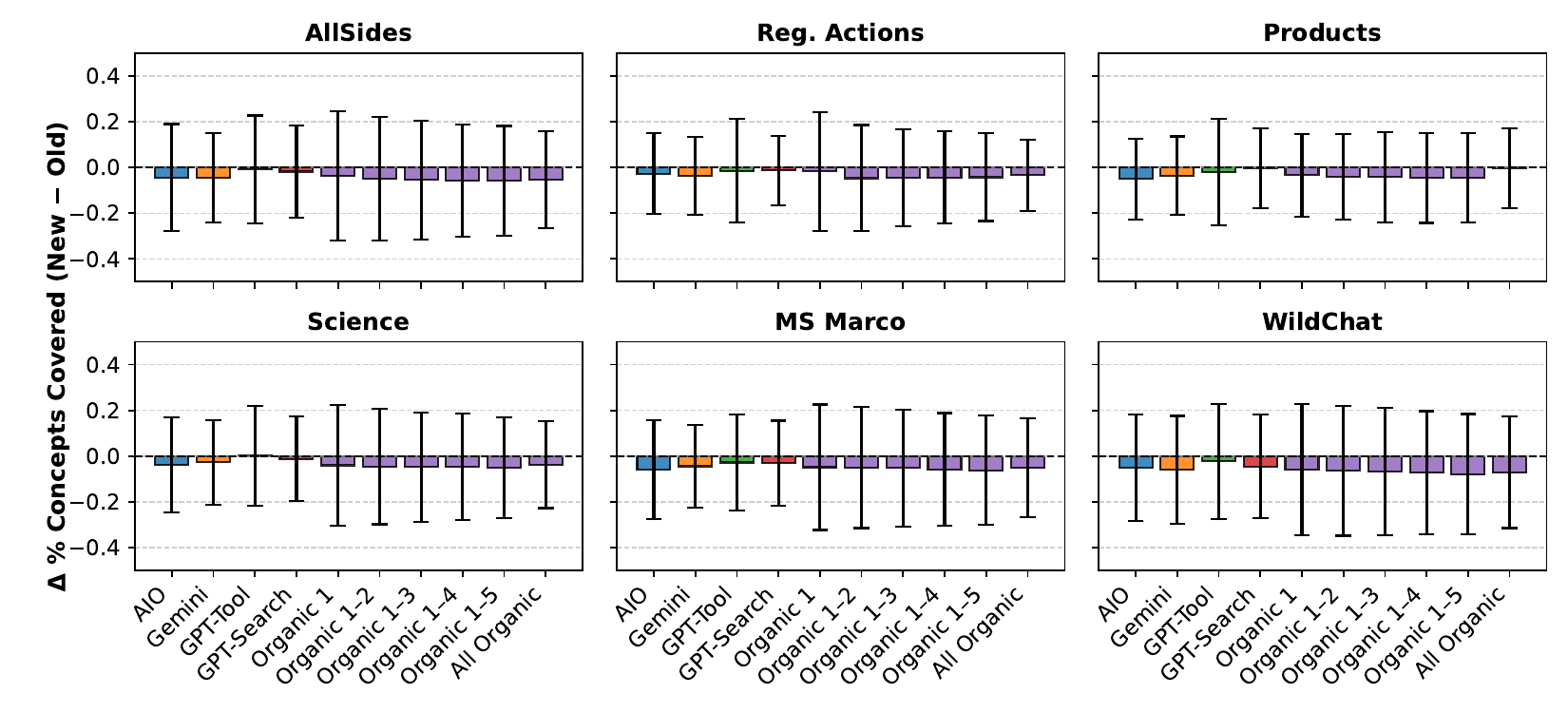}
        \caption{}
        \label{fig:newold_concepts}
    \end{subfigure}
    \caption{Left: Changes in average popularity rank per dataset between July/August and September 2025.
    \organic shows the smallest changes in ranks. Average ranks of generative engines change by as much as 40,000.
    Middle: Average link overlap per query between the two runs. \organic and \gemini are the most consistent.
    Right: Changes in average concept coverage per query.
    Changes in concept coverage are very small throughout.
    }
    \label{fig:newold}
\end{figure*}

Figure~\ref{fig:newold} shows the difference in rank distribution of retrieved web pages (Figure~\ref{fig:newold_ranks}) and the overlap between the links (Figure~\ref{fig:newold_overlap}) when comparing the July/August and September 2025 snapshots.
We do not include \sonar, as we only added the model in the September 2025 run.

\subsection{Details on Dataset Construction}
\label{app:yes_no_data}
We curate binary yes/no queries from the WildChat dataset~\citep{zhao2024wildchat1mchatgptinteraction} using a multi-stage filtering pipeline. First, we apply rule-based heuristics to identify interrogative prompts that are structurally compatible with binary responses, while excluding open-ended questions and those containing wh-words (e.g., what, which, who) that typically elicit non-binary answers. We then remove duplicate and near-duplicate queries to reduce redundancy. For the remaining candidates, we estimate the likelihood of a “yes” \vs “no” response using the Qwen3-30B-A3B-Instruct~\citep{qwen3technicalreport} model and retain only questions whose predicted probability of a “yes” response falls within a narrow ambiguity range of 0.4–0.6. This ambiguity-based criterion yields a balanced set of well-formed yes/no questions suitable for controlled comparisons across repeated executions. Finally, we perform a manual review to eliminate semantically trivial or meaningless queries (e.g., “can you speak Spanish”), ensuring the resulting dataset contains 209 substantive and informative binary questions.

\subsection{Details on LLM-based Judge}
\label{app:yes_no_judge}
We use an LLM-based judge to classify each output as either \textit{affirmative, negative, or mixed}. 
The model used is \texttt{gpt-5.2-2025-12-11} with Temperature 0.
The prompt used for classifying model responses:
\begin{tcolorbox}[
    breakable,
    fontupper=\footnotesize\ttfamily,
    left=1pt,
    right=1pt,
    top=1pt,
    bottom=1pt,
    standard jigsaw,
    opacityback=0
]
You will be given a question and a corresponding answer by a respondent.
\\
The question is designed to be answered in an affirmative (e.g., "yes") or negative (e.g., "no") manner. However, answers may be longer or phrased without explicit "yes" or "no".
\\
The answer can also contain both affirmative and negative elements, \eg depending on conditions or qualifiers. The answer can also express neither an affirmative nor negative stance, \eg due to uncertainty or avoidance. We call these cases "mixed".
\\
Your task is to classify the respondent's answer as "affirmative", "negative", or "mixed", and provide a brief rationale.
\\
\\
The question is:
\\
\{QUESTION\}
\\
The respondent's answer is:
\\
\{ANSWER\}
\end{tcolorbox}

The JSON response format used is shown in Listing \ref{lst:judge_schema}.

\begin{lstlisting}[language=json, caption={Judge response JSON schema}, label={lst:judge_schema}]
{
  "type": "object",
  "properties": {
    "stance": {
      "type": "string",
      "enum": ["affirmative", "negative", "mixed"]
    },
    "rationale": {
      "type": "string"
    }
  },
  "required": ["stance", "rationale"],
  "additionalProperties": false
}
\end{lstlisting}

\subsection{Details on Textual Overlap}
\label{app:textual_overlap}

Table~\ref{tab:jaccard_similarity} shows the textual overlap between repeated execution of the same queries at different timesteps (Section~\ref{sec:temporal_stability}).

\begin{table}[t]
\centering
\resizebox{\linewidth}{!}{
\begin{tabular}{lcccc}
\toprule
\textbf{Engine} 
& \multicolumn{2}{c}{\textbf{Temp. = 0}} 
& \multicolumn{2}{c}{\textbf{Default Temp.}} \\
\cmidrule(lr){2-3} \cmidrule(lr){4-5}
& $t_0 \rightarrow t_5$ 
& $t_0 \rightarrow t_{24}$
& $t_0 \rightarrow t_5$ 
& $t_0 \rightarrow t_{24}$ \\
\midrule
\gptnormal      & 0.63 $\pm$ 0.23 & 0.63 $\pm$ 0.23 & 0.31 $\pm$ 0.10 & 0.31 $\pm$ 0.10\\
\gptsearch      & 0.43 $\pm$ 0.16 & 0.42 $\pm$ 0.19 & 0.44 $\pm$ 0.19 & 0.41 $\pm$ 0.16 \\
\gemini         & 0.39 $\pm$ 0.14 & 0.38 $\pm$ 0.11 & 0.31 $\pm$ 0.09 & 0.32 $\pm$ 0.10 \\
\aio            & 0.47 $\pm$ 0.32 & 0.36 $\pm$ 0.22 & 0.50 $\pm$ 0.32 & 0.19 $\pm$ 0.21 \\
\sonar          & 0.27 $\pm$ 0.11 & 0.27 $\pm$ 0.11 & 0.34 $\pm$ 0.12 & 0.32 $\pm$ 0.11 \\
\bottomrule
\end{tabular}
}
\caption{
Mean Jaccard similarity of repeated executions.
Lower values indicate greater textual variability.
}
\label{tab:jaccard_similarity}
\end{table}

\section{Supplementary Material for \Cref{sec:knowledge_breadth}}
\label{app:link_analysis}

\subsection{Domain Category Classification}
\label{app:category_classification}

We classify URLs into various categories using their domain (\eg wikipedia.org).
We consider two different categorizations:
\begin{enumerate}
    \item {\bf Google Content Categories:} We use the 26 top-level categories from Google Content Categories.\footnote{\url{https://cloud.google.com/natural-language/docs/categories}} Example categories include ``Science'', ``News'' and ``Home and Garden''.
    \item {\bf Custom categories:} Google content categories can be too numerous and broad, \eg ``Games'', ``Reference''. We manually define the following categorization.
    \begin{enumerate}[i]
        \item News Media Site
        \item Science Publisher
        \item Encyclopedia
        \item Social Media \& User Forum
        \item Government
        \item NGO and Non-Profit
        \item Think Tank
        \item Corporate Entity
        \item Personal Blog
        \item None of the above
    \end{enumerate}
\end{enumerate}

We use \texttt{gpt4-turbo} to categorize each domain into the the categorization. The prompt we used is:

\begin{tcolorbox}[
    fontupper=\footnotesize\ttfamily,
    left=1pt,
    right=1pt,
    top=1pt,
    bottom=1pt,
    standard jigsaw,
    opacityback=0
]
Classify the following domain into one of the provided categories.
\\
Only respond with the name of the assigned category.
\\
\\
Domain: 'wikipedia.org'
\\
Categories:
\\
<list of categories>

\end{tcolorbox}

\subsection{Additional Information on \lloom}
\label{app:lloom}

We utilize \lloom \cite{Lam_2024}, an automated concept induction tool, to identify, score, and compare high-level concepts mentioned in the responses from different engines per query.

\lloom operates in two stages.
First, using a LLM, it analyzes a set of texts and proposes concepts of increasing generality, along with inclusion criteria and representative text examples.
In the second stage, it uses these concepts as labels to classify texts (again, by using a LLM), annotating if a given concept is present.%

\begin{figure}[t]
    \centering
    \includegraphics[width=\linewidth]{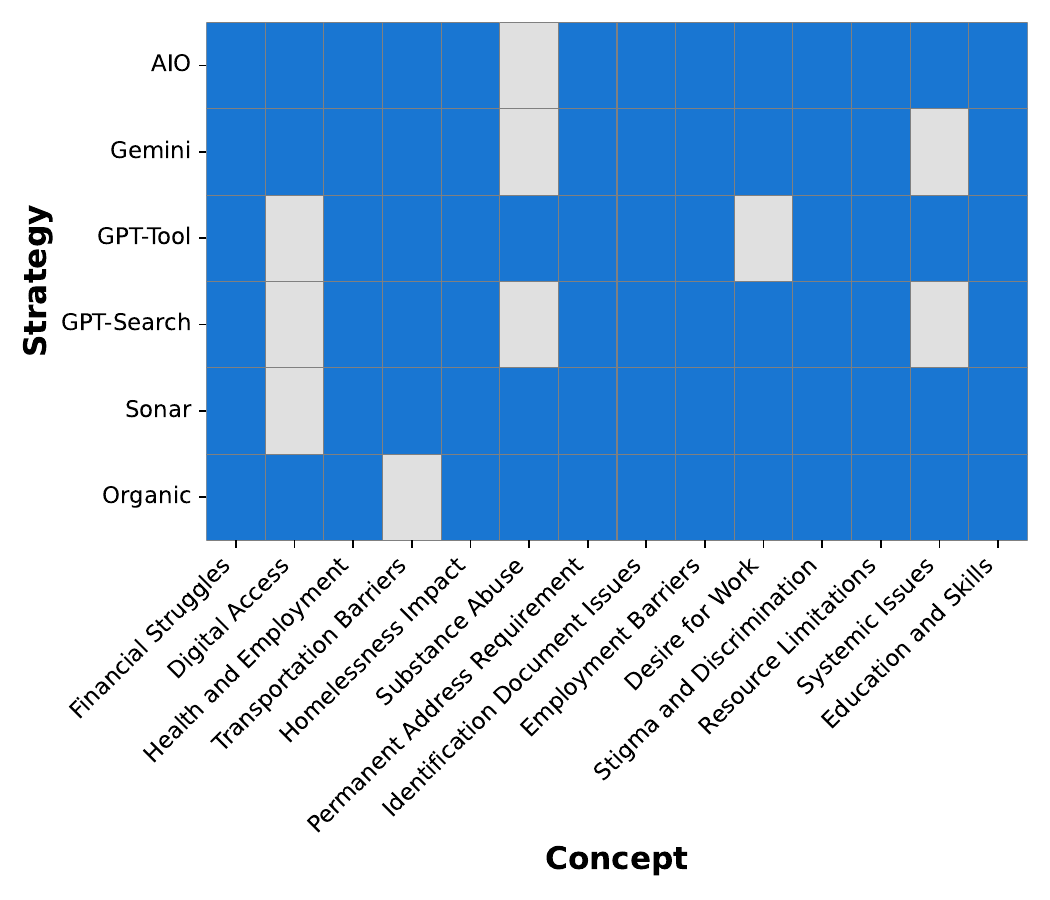}
    \caption{Concepts covered by different strategies for the query ``Why don't homeless people get jobs?'' (\allsides).
    The x-axis shows the topics discovered in the search engine outputs, the y-axis shows the search engines.
    Blue boxes indicate that the topic was present in the output of that search engine.
    Search engines differ in the concepts they bring up.}
    \label{fig:heatmap_example}
\end{figure}

For each query, we combine the output of all search engines, that is the four generative search engines and the \organic search, and apply \lloom to detect topics.
Each search engine's output is treated as a separate document except for \organic search. For \organic search, we treat each of the (usually 10) results as a separate document.
\lloom then produces a combined set of topics found in the collection of all the different engines' outputs for that query (stage 1).
We then classify each search engine's output against these topics to measure the fraction of identified concepts that each output contains (stage 2). \lloom classifies the presence of the topic at a 5-point scale of [``strongly disagree'', ``disagree'', ``neutral'', ``agree'', ``strongly agree'']. We deem a topic to be present if the classification output is ``agree'' or ``strongly agree''.
\Cref{fig:heatmap_example} shows an example of the topics surfaced for a query and how each engine scores them. Following the authors' implementation, we use \texttt{gpt-4o} for stage 1 and \texttt{gpt-4o-mini} for stage 2.

\begin{figure}[t]
    \centering
    \includegraphics[width=.9\linewidth]{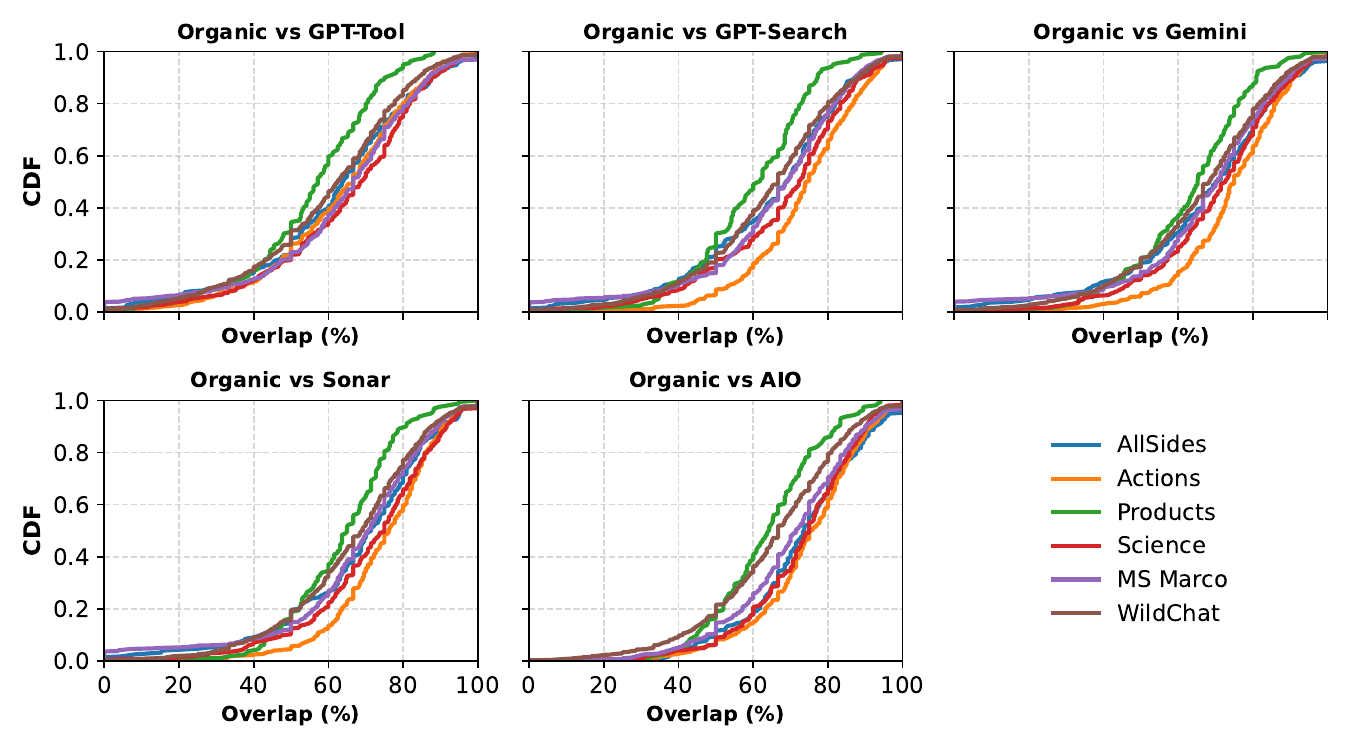}
    \caption{Overlap of concepts between \organic and generative search engines.
    }
    \label{fig:overlap_cdf_lloom_us}
\end{figure}

\begin{figure}[t]
    \centering
    \includegraphics[width=.9\linewidth]{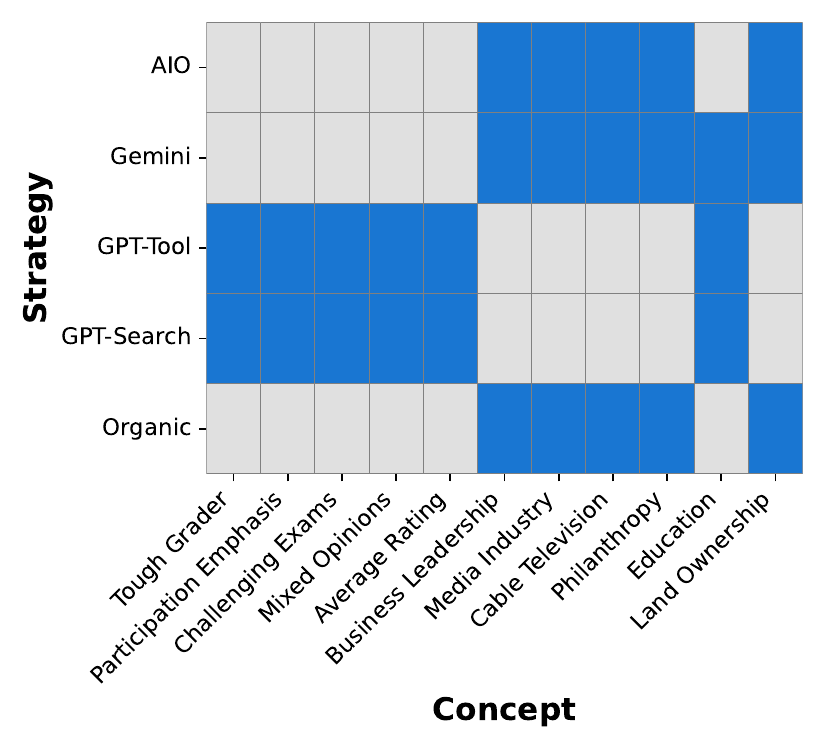}
    \caption{Topic coverage on the query ``who is john malone at cc'' (\wildchat).
    Blue boxes indicate that the topic was present in the output of that search engine.
    While \gptnormal and \gptsearch mention a professor called John Malone,  the other search engines refer to a billionaire businessman.}
    \label{fig:heatmap_query_0}
\end{figure}

\subsection{Remaining Figures from \Cref{sec:knowledge_breadth}}
\label{app:knowledge_breadth}

\begin{figure*}[ht]
    \begin{subfigure}[t]{0.19\linewidth}
        \centering
        \includegraphics[width=0.95\textwidth]{graphics/new_results/reach_100_bucket_us.pdf}
        \caption{\aio}
    \end{subfigure}%
    ~
    \begin{subfigure}[t]{0.19\linewidth}
        \centering
        \includegraphics[width=0.95\textwidth]{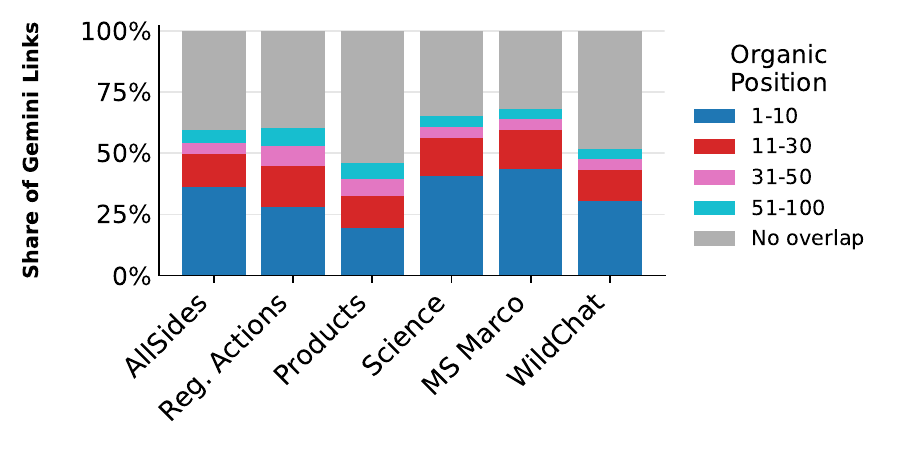}
        \caption{\gemini}
    \end{subfigure}%
    ~ 
    \begin{subfigure}[t]{0.19\linewidth}
        \centering
        \includegraphics[width=0.95\textwidth]{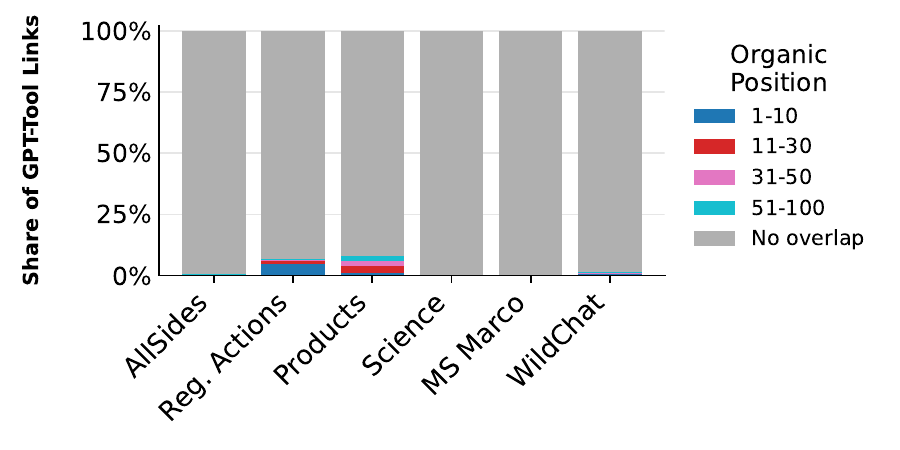}
        \caption{\gptnormal}
    \end{subfigure}
    ~ 
    \begin{subfigure}[t]{0.19\linewidth}
        \centering
        \includegraphics[width=0.95\textwidth]{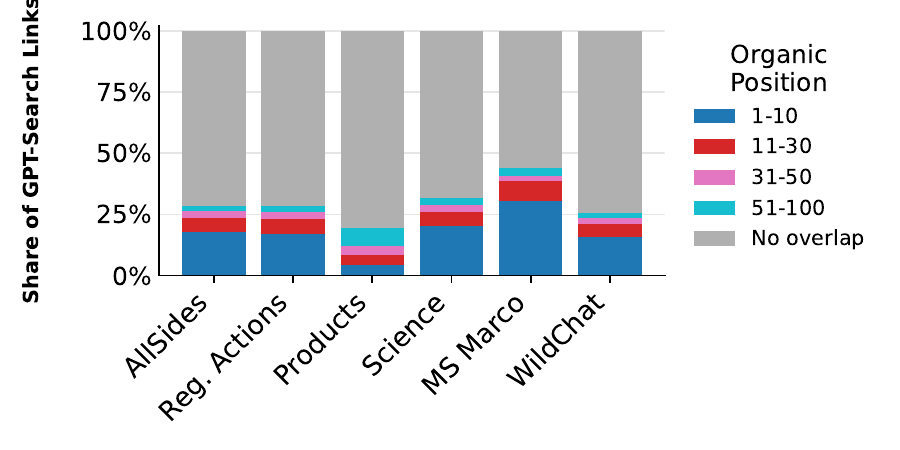}
        \caption{\gptsearch}
    \end{subfigure}
     ~ 
    \begin{subfigure}[t]{0.19\linewidth}
        \centering
        \includegraphics[width=0.95\textwidth]{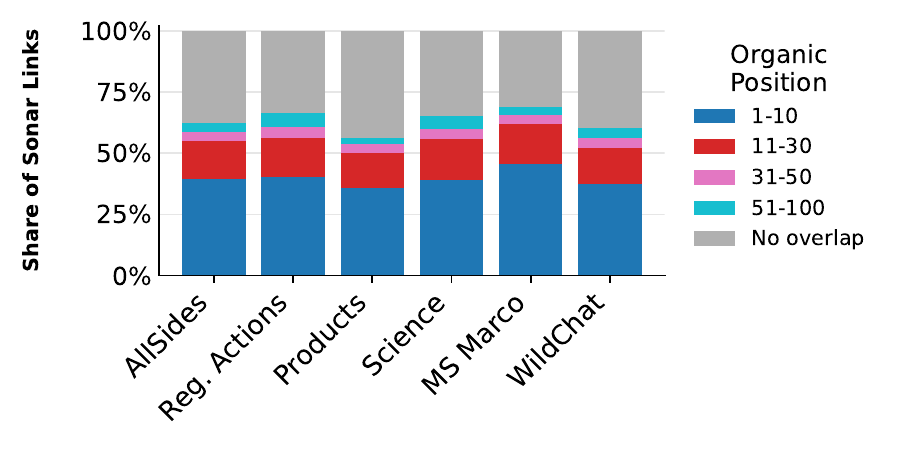}
        \caption{\sonar}
    \end{subfigure}
    \caption{
        [URL overlap] Overlap between the URLs retrieved by the top-100 \organic search results and the \aio,  \gemini, \gptnormal, \gptsearch, and \sonar models.}
    \label{fig:reach_100_stacked_gpts_gemini_us}
\end{figure*}

\begin{figure*}[ht]
    \begin{subfigure}[t]{0.19\linewidth}
        \centering
        \includegraphics[width=0.95\textwidth]{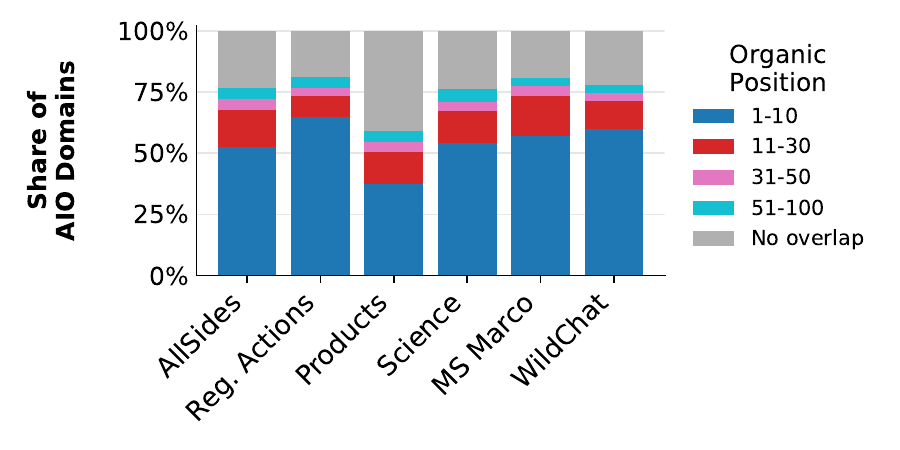}
        \caption{\aio}
    \end{subfigure}%
    ~
    \begin{subfigure}[t]{0.19\linewidth}
        \centering
        \includegraphics[width=0.95\textwidth]{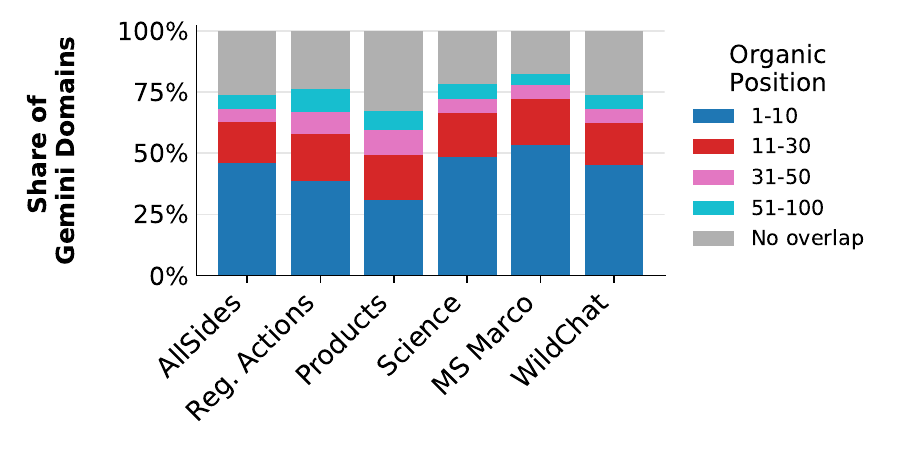}
        \caption{\gemini}
    \end{subfigure}%
    ~ 
    \begin{subfigure}[t]{0.19\linewidth}
        \centering
        \includegraphics[width=0.95\textwidth]{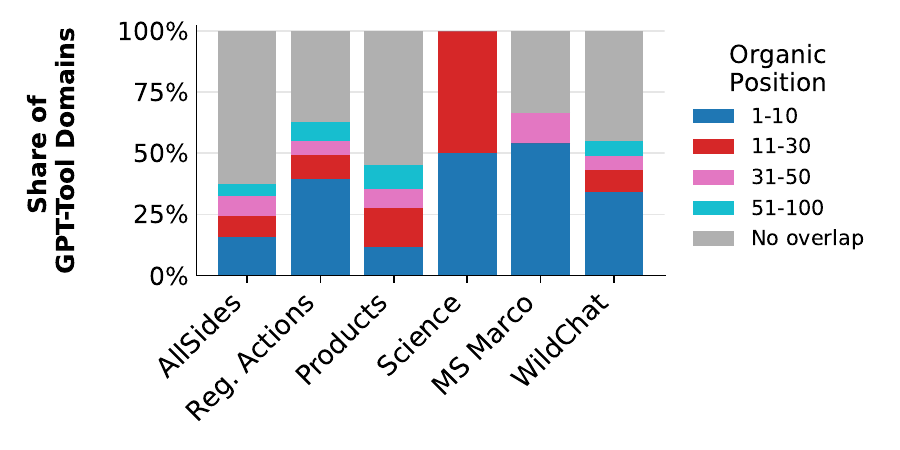}
        \caption{\gptnormal}
    \end{subfigure}
    ~ 
    \begin{subfigure}[t]{0.19\linewidth}
        \centering
        \includegraphics[width=0.95\textwidth]{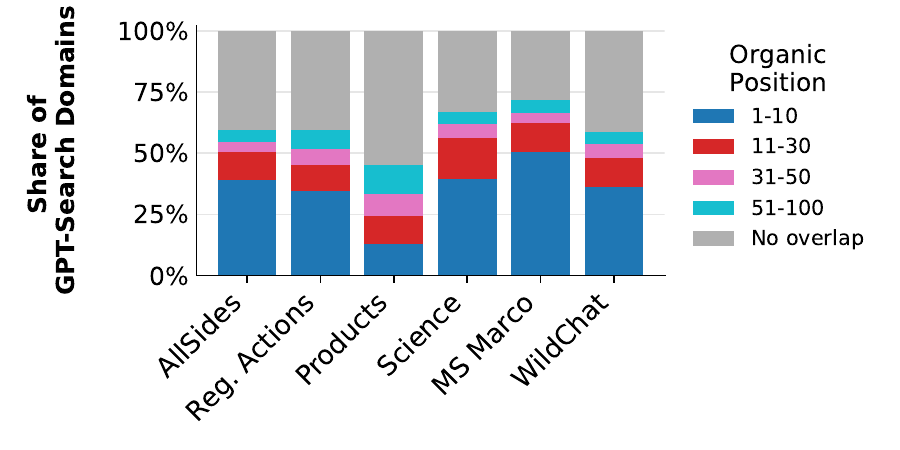}
        \caption{\gptsearch}
    \end{subfigure}
    ~ 
    \begin{subfigure}[t]{0.19\linewidth}
        \centering
        \includegraphics[width=0.95\textwidth]{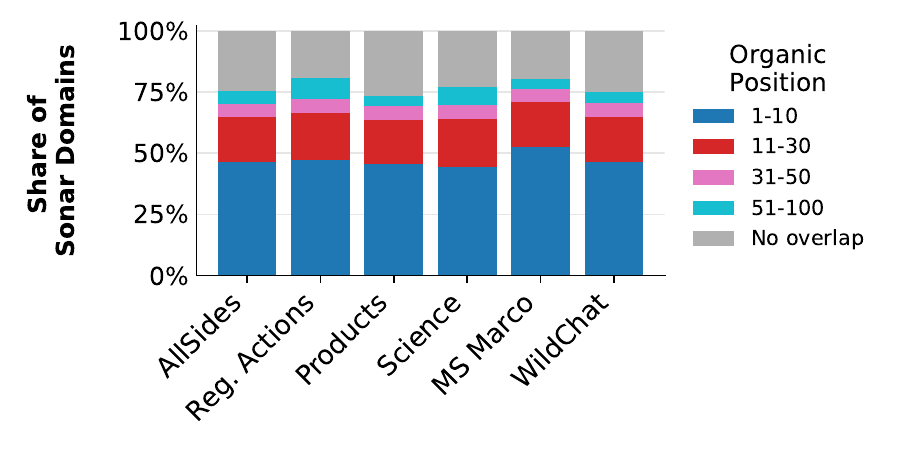}
        \caption{\sonar}
    \end{subfigure}
    \caption{
        [Domain overlap] Overlap between the domains retrieved by the top-100 \organic search results and the \aio, \gemini, \gptnormal, \gptsearch, and \sonar models.}
    \label{fig:reach_100_stacked_domain_gpts_gemini_us}
\end{figure*}

In this section, we report the overlap between the generative search engines and the top-100 links retrieved by \organic search.
Overlap increases when matching domains (see \Cref{fig:reach_100_stacked_gpts_gemini_us}) as compared to matching the URLs returned by the respective engines (see \Cref{fig:reach_100_stacked_domain_gpts_gemini_us}).
GPT models show the lowest overlap with organic search results overall, while \gemini had the highest overlap.
On \msmarco, \gemini captures more than $75\%$ of the top-100 links across queries.

\begin{figure*}[ht]
    \includegraphics[width=0.95\textwidth]{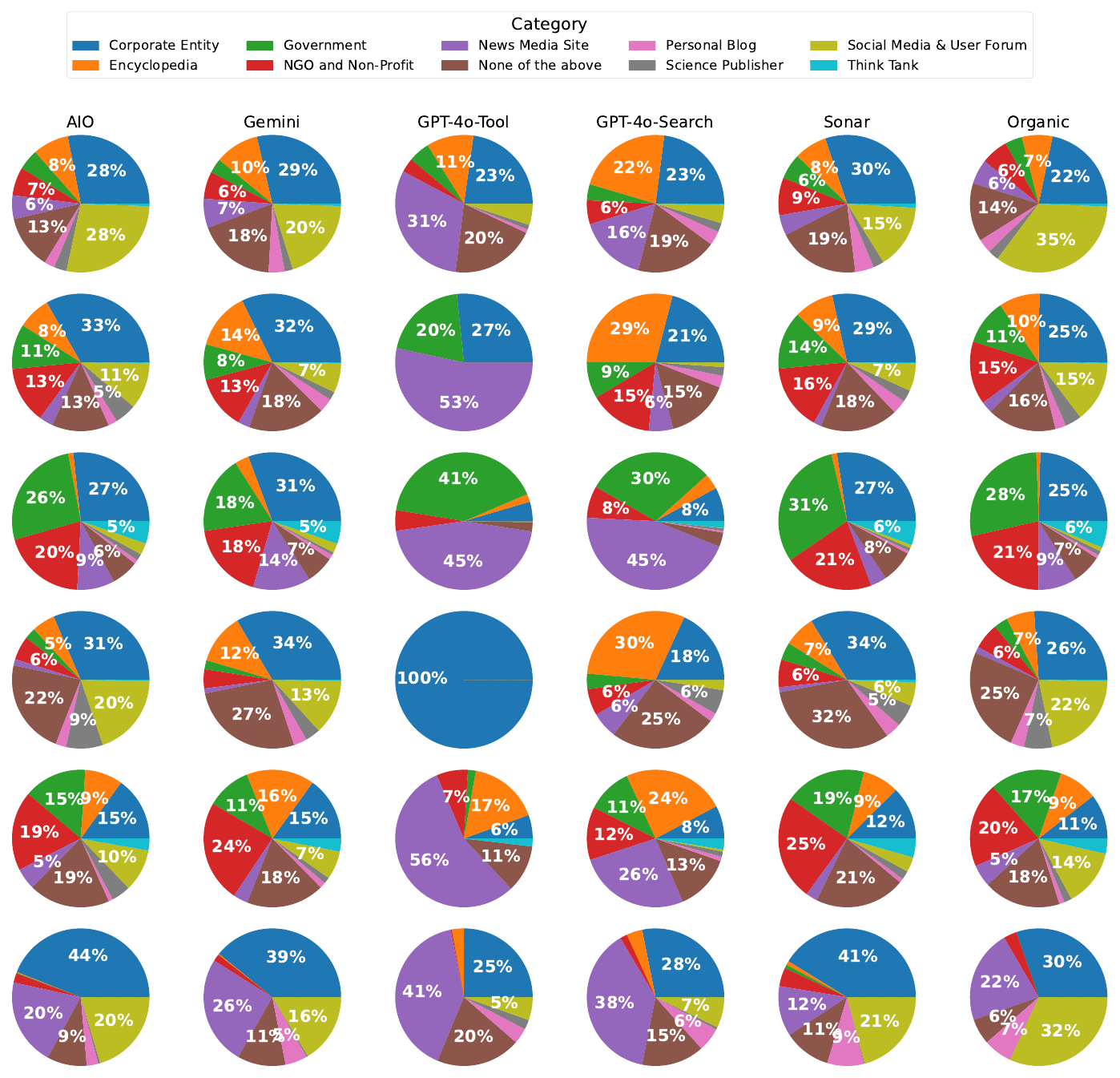}
    \caption{
    [Custom categorization] Categories of links retrieved by different search engines. The rows correspond to \wildchat, \msmarco, \regactions, \science, \allsides and \prods. For the same dataset, different search engines rely on different website types, \eg social media.
    }
    \label{fig:links_categories_us_custom}
\end{figure*}

\begin{figure*}[ht]
    \includegraphics[width=0.95\textwidth]{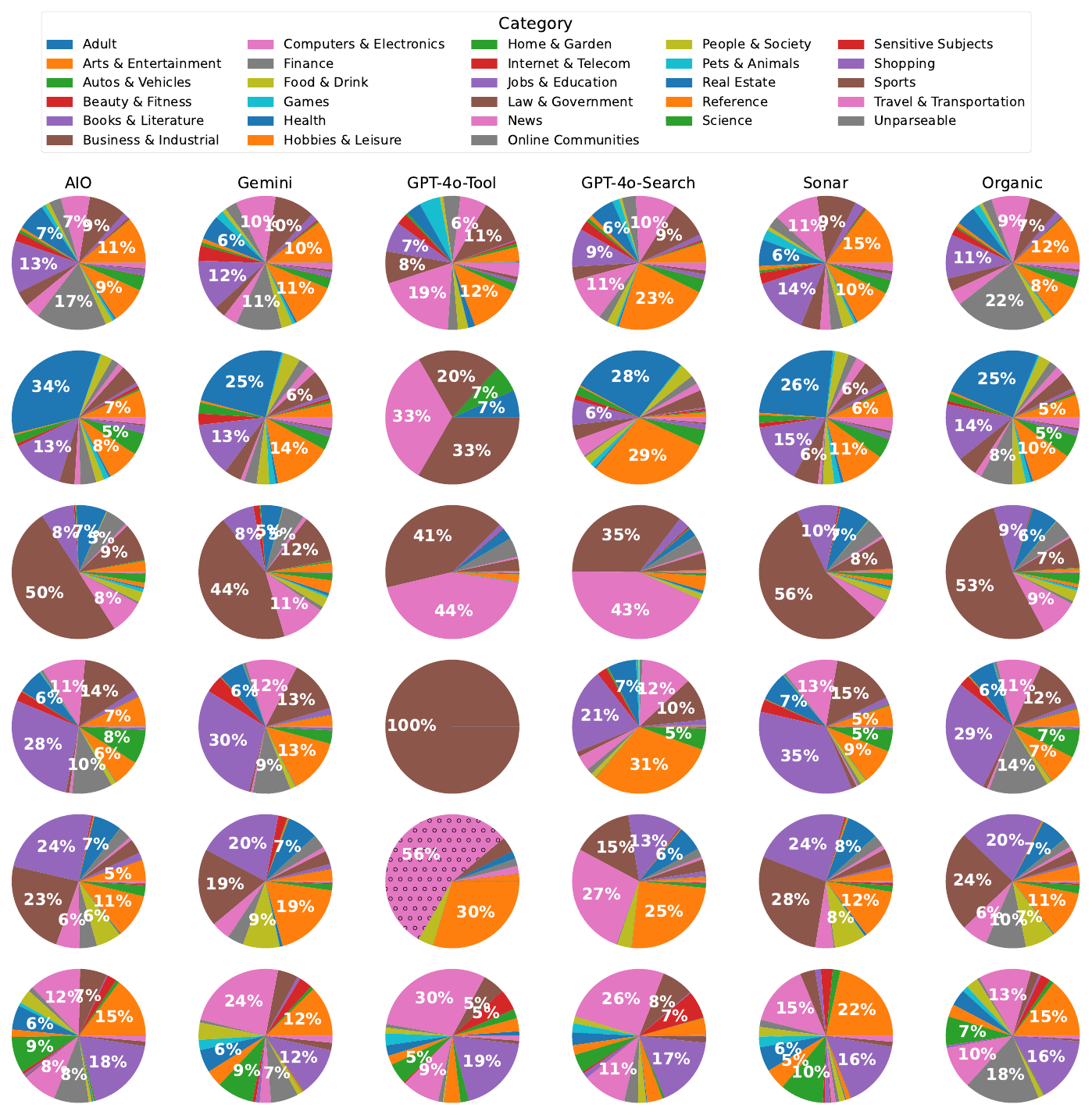}
    \caption{
    [Google content categorization] Categories of links retrieved by different search engines. The rows correspond to \wildchat, \msmarco, \regactions, \science, \allsides and \prods. For the same dataset, different search engines rely on different website types, \eg social media.
    }
    \label{fig:links_categories_us_google}
\end{figure*}

We also report the remaining pie chart analyses of the types of links retrieved by each engine for our custom categorization (\Cref{fig:links_categories_us_custom}) and Google's taxonomy (\Cref{fig:links_categories_us_google}).
For GPT models, we observe a high share of encyclopedias and news media websites. 
\organic search includes up to $35\%$ social media and user forums. However, few such domains are identified for GPT models.

\begin{figure}[t]
    \centering
        \includegraphics[width=.95\linewidth]{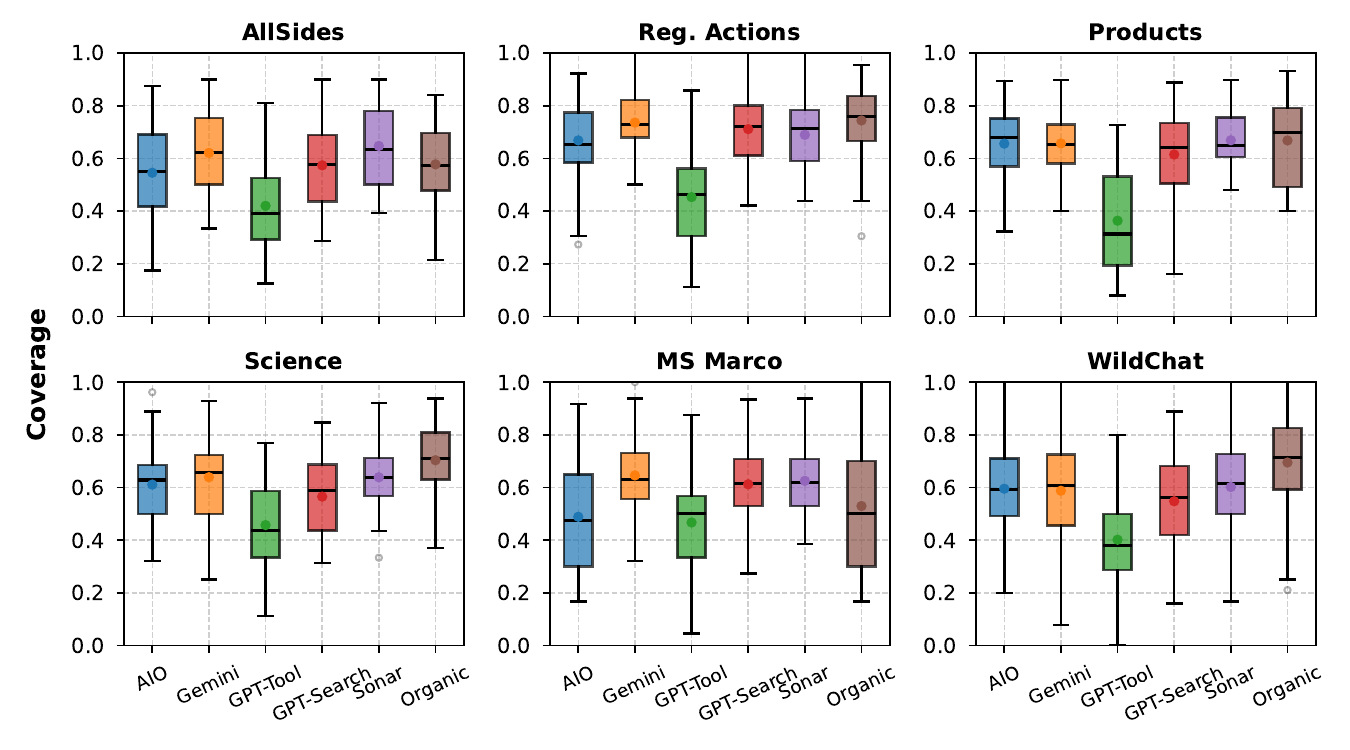}
\caption{Concept Coverage distribution on low coverage queries.
Low coverage queries are those where the fraction of concepts jointly covered by all strategies falls in the bottom 10th percentile of the distribution within a dataset.
\organic search maintains stable coverage on these queries, while generative engines like \gptnormal struggle to maintain good coverage.}
\label{fig:lloom_full_coverage_boxplot}
\end{figure}

\Cref{fig:lloom_full_coverage_boxplot} shows the concept coverage distribution on queries that have low concept coverage.

\section{Discussion of Results for the DE Location}
\label{app:de_analysis}

We replicate all experiments described in the main paper for queries issued from the DE (Germany) locale.
Overall, the qualitative trends and relative differences between search engines closely mirror those observed for the US location.
We do not observe systematic or statistically significant deviations in source diversity, topic coverage, or content characteristics.
The primary observable difference concerns the availability of Google AI Overviews.
Across datasets, AI Overviews are triggered less frequently in the DE locale than in the US (see Table \ref{table:data_stats}).
We leave further investigation of how geographic location influences generative search behavior to future work.

\section{Supplementary Analysis of Generative Search Outputs}
\label{app:supp_gensearch}

For GPT-based models, the API allows specifying a \textit{search context size} parameter with values \texttt{low}, \texttt{medium}, and \texttt{high}. 
This parameter controls the amount of external web information the model retrieves and integrates into its response, affecting cost, response quality, and latency.\footnote{\url{https://aiengineerguide.com/blog/openai-web-search-tool/}} 
Larger context sizes are expected to produce more comprehensive, but slower and more expensive, answers. 
Although this parameter has since been deprecated, we analyze its effects across the available settings (\texttt{low}, \texttt{medium}, \texttt{high}).

\xhdr{Minimal Impact of Search Context Size.}
Across all datasets, varying the search context size does not materially affect sourcing or response content.
The likelihood of performing a web search, the number of retrieved links per query, and the popularity rank of retrieved domains remain stable across context sizes.
Similarly, response length and topic coverage show no meaningful variation.
For example, \gptsearch achieves average concept coverage of $78\%$, $78\%$, and $77\%$ for \texttt{low}, \texttt{medium}, and \texttt{high} settings, respectively, while \gptnormal remains at $71\%$ throughout.
We also observe notable overlap in the concepts retrieved across different context sizes.
On \wildchat, for instance, the average concept overlap between any two context settings is approximately $65\%$.

\section{Supplementary Analysis of Webpage Content}
\label{app:supp_fullweb}
Our analysis of organic search is based on the top-10 results using titles, URLs, and snippets, rather than full webpage content. 
This choice ensures interface-level comparability, as generative engines present synthesized summaries rather than full documents.
Comparing to snippets also reflects typical user behavior, as prior work shows that users predominantly rely on the top-ranked results and often do not click through to full webpages~\cite{radlinskiQueryChainsLearning2006,10.5555/1484611.1484615,hochstotterWhatUsersSee2009}.

To examine whether this approximation understates the performance of traditional search, we conducted a pilot comparison on a small subset of queries ($n=60$) across datasets using full crawled webpage content.
We preprocess webpages by removing navigation elements and extracting the main content from HTML, excluding pages with fewer than $500$ characters to avoid incomplete or blocked content.
The $60$ queries have on average $7.4$ webpages available, with a minimum of $4$ pages per query.

\xhdr{Full webpages contain substantially more text.} Crawled web pages contain $26{,}983$ characters on average, compared to a maximum of $2{,}284$ characters for \gemini.
Running the \lloom analysis reveals that full webpages together cover on average $95\%$ of all concepts, compared to $50{-}67\%$ across generative systems.
These results show that considering the full web page content leads to a \textbf{higher recall than any generative engine}.
However, this higher recall comes at the cost of \textbf{lower precision}:  
On average, $21\%$ of all concepts are only covered by webpages and not by any other engine, suggesting that websites contain significant amounts of additional information that is often only loosely related to the query.
Due to the length of the webpages, \textbf{some concepts can be unrelated to the user query}, \eg for a query about when Tetris was created, some of the concepts that are only surfaced by full webpages are ``Future Gaming Trends'', ''Cold War Context``, and ``Scientific Influence''.
This recall–precision trade-off highlights a structural distinction between paradigms: organic search exposes broader document content, whereas generative systems prioritize concise, synthesized summaries.

\end{document}